\documentclass[a4paper,openany, 12pt]{book}

\usepackage[text={15.5cm,23cm},centering]{geometry}
\usepackage[sectionbib]{chapterbib} 
\usepackage{mathptmx}
\usepackage{graphicx}
\usepackage{subfigure}
\usepackage{color}
\usepackage[colorlinks=true, linkcolor=blue]{hyperref}
\usepackage{natbib}
\usepackage{deluxetable}
\usepackage{amssymb}
\usepackage{aas_macros}
\usepackage{longtable} 

\newcommand{\chapterauthor}[1]{\textsc{#1}\section*{}} 

\begin{document}
\setcounter{chapter}{8}

\chapter{The Near-Centaur Environment: Satellites, Rings, and Debris}

\chapterauthor{A.A.~Sickafoose$^1$, S.M.~Giuliatti Winter$^2$, R.~Leiva$^3$, C.B.~Olkin$^4$, D.~Ragozzine$^5$, L.~Woodney$^6$}
$^1$ Planetary Science Institute, 1700 E. Fort Lowell, Suite 106, Tucson, AZ 85719, U.S.A.\\
$^2$ Grupo de Din\~amica Orbital \& Planetologia, S\~ao Paulo State University - UNESP, Av. Ariberto Pereira da Cunha, 333, Guaratinguet\'a, SP 12516-410, Brazil\\
$^3$ Instituto de Astrof\'isica de Andalucía, – Consejo Superior de Investigaciones Cient\'ificas (IAA-CSIC), Glorieta de la Astronom\'ia S/N, E-18008 Granada, Spain\\
$^4$ Muon Space, 2250 Charleston Rd., Mountain View, CA 94043, U.S.A.\\
$^5$ Brigham Young University Department of Physics \& Astronomy, N283 ESC, Brigham Young University, Provo, UT 84602, U.S.A.\\
$^6$ Department of Physics, Cal State University San Bernardino, 5500 University Parkway, San Bernardino, CA 92407, U.S.A.\\




\newcommand{\silvia}[1]{{\color[rgb]{0.0, 0.5, 1.0} {\bf Silvia:} #1}}
\newcommand{\laura}[1]{{\color[rgb]{0.3, 0.3, 0.3} {\bf Laura:} #1}}
\newcommand{\rodrigo}[1]{{\color[rgb]{0.0, 0.0, 1.0} {\bf Rodrigo:} #1}}
\newcommand{\amanda}[1]{{\color[rgb]{0., .5,0.} {\bf Amanda:} #1}}
\newcommand{\cathy}[1]{{\color[rgb]{0.5, 0,0.5} {\bf Cathy:} #1}}
\newcommand{\darin}[1]{{\color[rgb]{0.7, 0.3,1} {\bf Darin:} #1}}

\textbf{Abstract:} The unexpected finding of a ring system around the Centaur (10199) Chariklo opened a new window for dynamical studies and posed many questions about the formation and evolutionary mechanisms as well as the relation to satellites and outbursting activity. As minor planets that cross the orbits of the giant planets, Centaurs have short dynamical lifetimes: Centaurs are supplied from the Trans-Neptunian region and some fraction migrates to become Jupiter-family comets (JFCs). Given these dynamical pathways, a comparison of attributes across these classifications provides information to understand the source population(s) and the processes that have affected these minor planets throughout their lifetimes. In this chapter we review the current knowledge of satellites, rings, and debris around Centaur-like bodies, discuss the observational techniques involved, place the information into context with the Trans-Neptunian Objects (TNOs), and consider what the results tell us about the outer Solar System. We also examine open questions and future prospects.
\section{Introduction}\label{sec:Intro}
This chapter focuses on the presence or lack thereof of satellites, rings, and/or surrounding debris found around bodies in the Centaur region. Table \ref{tab:srd_stats} contains all the Centaurs and closely-related objects for which these features have been detected.
\subsection{Satellites, Rings, and Debris in the Centaur Region}\label{subsec:SRD}
There are no known Centaur binaries using the definition of objects with perihelion distances $q>5.2 \, {\rm AU}$ and heliocentric semimajor axes $a_\odot<30  \, {\rm AU}$. However, there are two related objects that are interesting to consider: these are Giant Planet Crossers (GPCs), with perihelion distances $5.2<q<30  \, {\rm AU}$ (see Chapter 3). The two known GPC binaries are (42355) Typhon-Echidna and (65489) Ceto-Phorcys.
 
\begin{deluxetable}{lcccccl}
\tablecaption{Published Detections of Satellites, Rings, and Debris around Centaurs (top) and Giant Planet Crossers (GPCs; bottom)\textsuperscript{a}.\label{tab:srd_stats}}
\tablewidth{0pt}
\tablehead{
\colhead{Body}&\colhead{$q$; $a_\odot$}& \colhead{Satellites\textsuperscript{c}} &\colhead{Rings\tablenotemark{c}} & \colhead{Debris\textsuperscript{c}}& \colhead{Type\textsuperscript{d}} & \colhead{Ref.} \\
\colhead{}&\colhead{(AU)\textsuperscript{b}}& \colhead{} &\colhead{} & \colhead{}& \colhead{} & \colhead{}
}
\startdata
(2060) Chiron & 8.6; 13.7 &-\textsuperscript{e} & Maybe\textsuperscript{f} & Yes & I:O:IR & \citep{Bus1996,Elliot1995,Ruprecht2015}\textsuperscript{g} \\
 & & & & & & \cite{Ortiz2015,Ortiz2023,Sickafoose2023} \\
(10199) Chariklo &13.1; 15.8& -\textsuperscript{e} & Yes & - & O;IR & \cite{Braga-Ribas2014,Leiva2017,Berard2017}\textsuperscript{g} \\
 & & & &  &  & \cite{Morgado2021,Santos-Sanz2023} \\
(60558) Echeclus &5.8; 10.8& -\textsuperscript{e} & No\textsuperscript{h} & Yes & I;O;IR& \citep{Rousselot2016,Bauer2008,Pereira2024b}\textsuperscript{g} \\
29P/Schwassmann- &5.8; 6.0& -\textsuperscript{e} & No\textsuperscript{h} & Yes & I;O;IR & \cite{Jewitt1990,Wierzchos2020}\textsuperscript{g}\\
Wachmann 1 (SW1)&& & & & & \cite{Buie2023} \\
\hline
(42355) Typhon &17.5; 37.5& Binary & - & - & I & \cite{Noll2006,Grundy2008,Stansberry2012} \\
(65489) Ceto &17.7; 99.2& Binary & - & - & I & \cite{Grundy2007}\\
\enddata
\tablecomments{\textsuperscript{a}{Where Centaurs are defined as $q>5.2\, {\rm AU}$ and $a_\odot < 30 \, {\rm AU}$ and GPCs have $5.2 \, {\rm AU} < q < 30 \, {\rm AU}$.} 
\textsuperscript{b}{Perihelion distance, $q$, and heliocentric semimajor axis, $a_\odot$. Most objects are detected near perihelion.} 
\textsuperscript{c}{Satellites, rings, and debris have neither been detected nor ruled out for objects with no entry in these columns.} 
\textsuperscript{d}{Observation type: I – imaging (visible wavelength), O – stellar occultation, IR - infrared imaging or spectroscopy.}
\textsuperscript{e}{HST has observed these objects and would be sensitive to $\sim$km-sized satellites at separations of $\gtrsim$0.5 arcseconds (typically $\gtrsim$7000 km) and larger satellites at closer separations. For example, at $\sim$700 km, an equal-brightness binary could be detected.}
\textsuperscript{f}{For Chiron, secondary detections in stellar occultations have been interpreted as a jet \citep{Bus1996,Elliot1995}, a shell or arcs \citep{Ruprecht2015}, a two-ring system \citep{Ortiz2015}, and most recently as changing material \citep{Ortiz2023,Sickafoose2023}.} 
\textsuperscript{g}{Also references therein: these targets have been extensively observed, as reported in Chapters 10 and 11.}
\textsuperscript{h}{Stellar occultations were used to place upper limits on ring material for Echeclus, and initial light curves did not show signs of ring-like dips for SW1 (see Section \ref{subsec:ringchars} \cite{Pereira2024b,Buie2023}).} 
}
\vspace{-30pt}
\end{deluxetable}

The number of Centaurs that are known to have rings is likewise low, at one or possibly two objects (see Table \ref{tab:srd_stats}): Chariklo (rings dubbed C1R and C2R) and (2060) Chiron. The ringed objects are two of the largest Centaurs, with effective diameters $\gtrsim200 \, {\rm km}$.  The first discovery of a small-body ring system is relatively recent, via a stellar occultation in 2013 \citep{Braga-Ribas2014}. The success rate of this technique has benefited from advances in astrometric measurements as well as instrumentation and larger campaigns (see Chapter 11). Notably, the subsequent discovery of ring material around the two large TNOs (136108) Haumea and (50000) Quaoar \citep{Ortiz2017,Morgado2023} prompts consideration of whether rings form in the Centaur region or survive the transition from the Trans-Neptunian region \citep[e.g. ][]{Araujo2016}.

Finally, while there are tens of known active Centaurs (e.g. Chapter 8 and \citep{Chandler2020}), only three Centaurs have had observable surrounding debris (Table \ref{tab:srd_stats} and references therein): Chiron, (60558) Echeclus, and SW1. We consider debris to be material consisting of larger particles than the gas and dust typically released from comets, which could have significant optical depth to be detected in stellar occultations. Fewer than a dozen Centaurs have been successfully observed using occultations, leaving ample space for more discoveries of rings and debris. Note that many comets could qualify as GPCs surrounded by debris. For the purposes of this chapter, we consider only the GPC binaries and the GPCs for which stellar occultations have been reported.

More details are provided about the known satellites, rings, and debris of Centaur-like bodies in Sections \ref{sec:sats}-\ref{sec:debris}.

\subsection{Overview of Observing Techniques}\label{subsec:techniques}

Satellites, rings, and debris can be or have been detected around Centaur-like objects through both direct and indirect methods, including occultations, imaging, spectroscopy, and photometry-generated light curves. The applications of these techniques are discussed briefly in this section.

\subsubsection{Direct-detection methods} \label{subsubsec:directtechs}

\textit{Stellar occultations:} Six Centaurs have published stellar occultation results: Chiron \cite[e.g.][]{Bus1996}, (8405) Asbolus \cite{Rommel2020}, (54598) Bienor \citep{Fernandez-Valenzuela2023,Fernandez-Valenzuela2017}, Chariklo \cite[e.g.][]{Braga-Ribas2014}, Echeclus \cite{Pereira2024b}, and (95626) 2002 GZ\textsubscript{32}\cite{Santos-Sanz2021} (also see references in Table \ref{tab:srd_stats}). Additionally, stellar occultations have been reported for Centaur 2008 YB\textsubscript{3}, the GPC 2014 YY\textsubscript{49}, and the Resonant TNO (591376) 2013 NL\textsubscript{24} \cite{Strauss2021}, the latter of which meets our criteria for being a GPC. 

\begin{figure}[bt]
\centering
\includegraphics[width=1\textwidth,clip,trim=0mm 0mm 0mm 0mm]{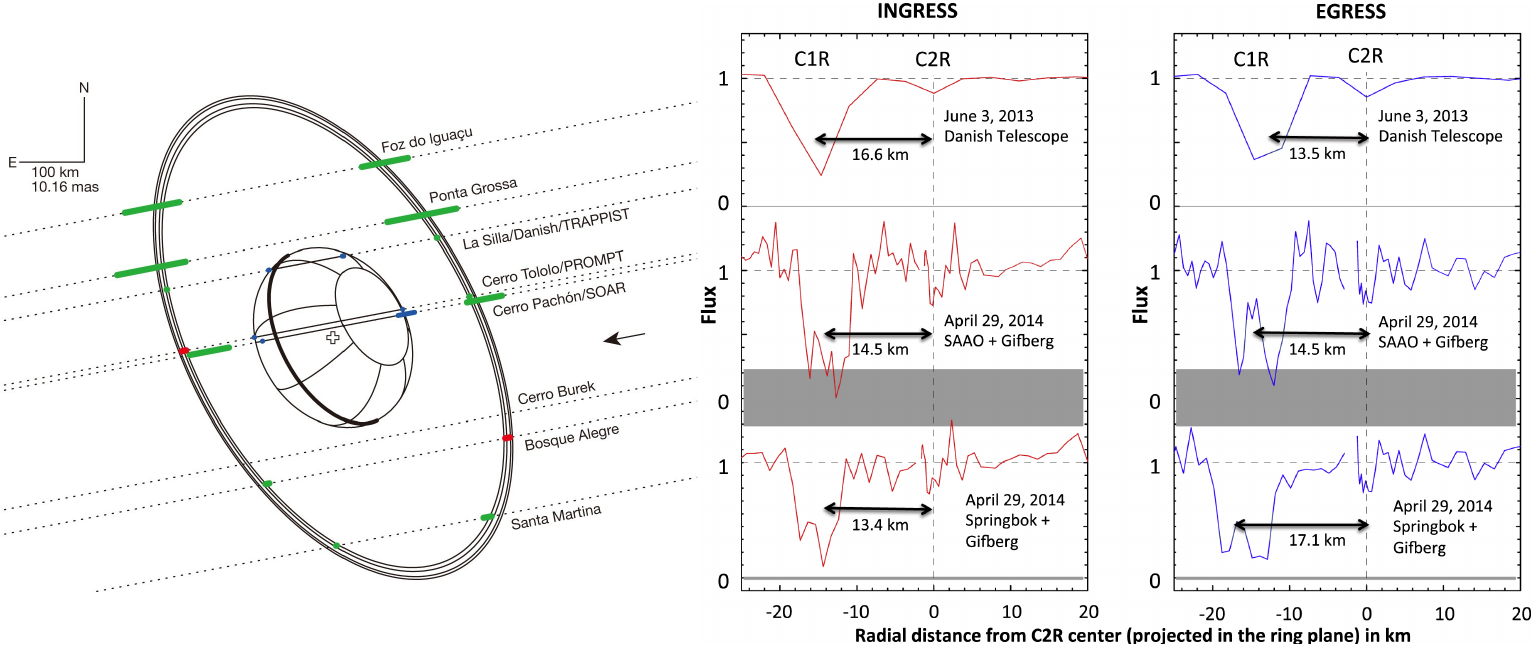}
\caption{Example of ring detections from stellar occultation data. (\textit{left}) Sky-plane view of Chariklo with occultation chords in 2013 (Fig.~2 in \cite{Braga-Ribas2014}). The green segments show locations of ring detections, with the length of the segments representing the uncertainties (red represents ring locations during camera readouts). Blue segments detecting the central object allowed constraint of the geometry of the rings. (\textit{right}) Radial profiles of Chariklo's rings during stellar occultations (Fig.~7 in \cite{Berard2017}). The flux of the star plus Chariklo is normalized to unity, and dips indicate ring material blocking the stellar flux. The ring profiles at the time of discovery are shown in the top panel for the inner and outer rings, named C1R and C2R, respectively, as a function of radial distance. The lower panels show more detailed structure a year later, indicating longitudinal variations in the distance between the rings.
\label{fig:occ}}
\end{figure}

Stellar occultations are a well-established method for detecting and characterizing planetary rings \citep[e.g. ][]{Elliot1978, Bosh2002}. In a stellar occultation, the planetary body is observed passing in front of a distant star. The starlight acts as a probe of the body and its vicinity. Extinction of the starlight due to rings or debris can provide astrometric information of the location, as well as an indication of the amount of surrounding material (see Fig. \ref{fig:occ}). In order to derive accurate relative positions, an occultation of the rings/debris and primary body are both needed. This can be accomplished with one observing site or with separate observations of each component by different observers, given that separate sites have precise timing and well-known coordinates for the observing locations. Today, multiple-site scenarios are enabled by global navigation satellite systems that provide accurate position and timing services.

Stellar occultations are recorded with a series of images, preferably accurately timed. Each image in the sequence provides a unique probe of the planetary system as the apparent location of the star moves relative to the body. High-speed imaging allows derivation of spatial information on a scale that is smaller than the diffraction limit of direct-imaging systems. The cadence of observations is driven by the required signal-to-noise ratio (SNR) of the data, which depends on the science objective of the observations. There is a trade between achieving high SNR and achieving high spatial sampling. Higher SNRs can be achieved without compromising the spatial sampling of the data for brighter stars, lower shadow velocities, and/or larger telescope apertures.

The observed extinction due to ring, dust, or atmospheric particles depends on the wavelength of light being observed, particularly for particles of the same size or smaller than this wavelength. For this reason, additional information may be retrieved from a stellar occultation if multiple wavelengths are used in the observations. From the reported occultation results, only two Centaurs are thought to possibly have rings (Chariklo and Chiron), and Chariklo-like rings have been ruled out for two other Centaurs (Echeclus and SW1). Observations with higher SNR and/or spatial sampling are still needed to detect or rule out rings like those around Chariklo at all other Centaurs and GPCs. 

Stellar occultations by comets are generally more difficult to observe than stellar occultations by Centaurs and TNOs. Complications include (i) the prediction of the shadow path, because of non-gravitational forces and the difficulty in detecting the nucleus of an active comet \citep{Miles2018}, and (ii) the typically small size of the nucleus, making the shadow path on the Earth narrow \citep{Combes1983,Miles2018}. 
The difficulty of using occultations to characterize debris around outbursting objects is demonstrated by reported stellar appulses for comets SW1, Hale-Bopp, and 17P/Holmes  \citep{Fernandez1999,Lacerda2012, Miles2021}. The stars passed near, but not behind, the nuclei during these events. An optically thick coma was only detected within 100 km of Hale-Bopp's nucleus and no other significant debris was observed in the occultation data.

Multichord occultations are useful for determining sizes and shapes of small bodies. Currently, only Chariklo, Chiron, and Echeclus have sufficient occultation data to determine well-constrained, triaxial shapes (see Section \ref{subsec:ringchars}, Chapter 4, and \cite{Pereira2024b}). For the other objects, upper limits on diameters and 
axis ratios have been placed. Stellar occultations can additionally be used to discover satellites, such as occurred for Neptune's moon Larissa \cite{Reitsema1982}. However, the odds are low due to the typically small sizes of any moons and the need for favorable geometry. Stellar occultations are more likely to be successful for a satellite once its orbit is well established, as has been done for the TNOs Quaoar (Weywot \cite{Fernandez-Valenzuela2023b}) and Orcus (Vanth \cite{Sickafoose2019}).

\textit{Imaging:} Direct imaging is the most productive method to discover and characterize binary objects. The sensitivity to detecting companions depends on factors such as the components' angular separation, position angle, and brightness ratio, translating into significant variations in the satellite detection limit among the objects observed so far. When the two objects in a binary are separated by at least one Point Spread Function Full Width at Half Maximum (PSF FWHM), then they can typically be resolved as more than one object, especially if they are near equal brightness. Observations with a stable and small PSF are thus desirable to find binaries as close as possible, which has led to the extensive use of the Hubble Space Telescope (HST) for direct imaging campaigns. Indeed, all the objects in Table \ref{tab:srd_stats} have been observed by HST at least once. HST can often resolve near-equal-brightness binary components that are separated by $\gtrsim0.05 \, {\rm arcseconds}$, which translates to $\sim$700~km at 20~AU. Faint binary components are harder to detect close-in; however, Centaur satellites a few kilometers in size could potentially be detected at wide separations of several thousand kilometers. Whether multiple components can be discovered depends on the separation at the time of the observation: even orbits much larger than the resolution limit cannot be discarded, because the observations could have occurred when the two objects happened to be aligned with respect to the observer. This is more common for edge-on orbits observed only once and might be avoided entirely for more face-on orbits.

\begin{figure}[ht!]
\centering
\includegraphics[width=1\textwidth,clip,trim=0mm 0mm 0mm 0mm]{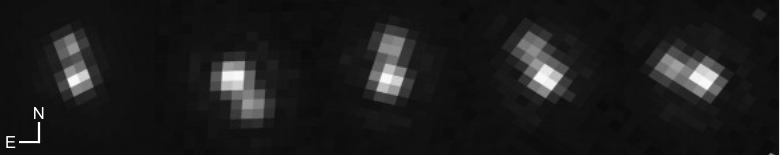}
\caption{The binary GPC Ceto-Phorcys observed by HST (Fig. 2 in \cite{Grundy2007}). The mosaic is 2.0 by 0.4 arcsec, with each frame centered on Ceto. From left to right, the images were taken between 2006 April (discovery) and May (followup).
\label{fig:imaging_1}}
\end{figure}

Typically, upon discovery of a binary, more HST time is requested and granted to make multiple observations in order to fully determine the parameters of the mutual orbit. For example, in \cite{Grundy2007}, five HST observations of Ceto and its companion Phorcys allowed for the measurement of the period, semimajor axis, eccentricity, and total system mass (see Fig. \ref{fig:imaging_1}). In this case, a ``mirror degeneracy" remains so that there are two possible orientations for the system in three-dimensional space. The standard has been to assume a Keplerian orbit, but new methods now allow for the exploration of non-Keplerian effects \cite[e.g.][]{Ragozzine2024, Proudfoot2024}. These non-Keplerian effects are most typically caused by non-spherical shapes or unknown components. Modeling can thus provide additional information on shapes if there are sufficient observational constraints over a long enough time to detect the slow orbital changes due to non-Keplerian effects. 

Planetary rings are not easily detectable in images because of the high contrast with the nearby parent body. Visible-wavelength imaging has been unsuccessful to date for observing Centaur rings, even with HST for the known system at Chariklo \citep{Berard2017}. However, imaging of Centaur rings from space might be feasible at some point. Dusty rings have been observed by the \textit{Spitzer} Space Telescope \cite[albeit on a massive scale at Saturn,][]{Verbiscer2009} and JWST has promising capabilities for studying ring systems \cite{Tiscareno2016}. Currently, direct detection of Chariklo's rings is being investigated by subtracting out the signal from the nucleus for recent JWST observations \cite{Santos-Sanz2023}. Such measurements would provide a direct way to study the evolution of rings at higher temporal resolution than the currently-available method (stellar occultations).

Imaging at different wavelengths is currently the primary method to identify and characterize debris around Centaurs. Outbursting activity on Centaurs has been detected in visible-wavelength images (see Chapters 8 \& 10). At these wavelengths, sunlight is scattered off of dust particles (as well as some ions and radicals) in the comae \cite{Cochran1991}. Dust and gas production rates can be determined by studying the brightness profiles of extended objects, where the PSFs do not match the field stars \citep[e.g. ][]{A'hearn1984,Fink2021}. Changing color indices over time can further indicate varying dust-size distributions \citep[e.g. ][]{Rousselot2008}. Infrared images and spectra can be used to study morphology, constrain particle sizes, and determine dust composition \citep[e.g. ][]{Bauer2008,Stansberry2004}. Furthermore, submillimeter data combined with visible can be used to look for correlations between gas and dust outbursting \citep[e.g. for CO; ][]{Wierzchos2020}. The detection of bigger debris in images would require searching for any separated components, like those in the break-up of active asteroid P/2013 R3 \cite{Jewitt2014}. An example analysis for ejected material is of a ``secondary" source that was observed at Echeclus after outbursting, which was determined to not be an ejected fragment but rather the result of localized cometary activity \cite{Weissman2006,Bauer2008,Rousselot2016}.

\textit{Spacecraft:} The best way to directly detect small satellites and rings around bodies in the outer Solar System would be to send a spacecraft \cite[e.g.]{Singer2021}. However, the cost and long timelines for spacecraft missions are often prohibitive. In addition, the debris environment must be sufficiently characterized to ensure a low risk of spacecraft damage \citep[e.g. ][]{Fink2021}. 

\subsubsection{Indirect Methods}\label{subsubsection:indirecttechs}
\textit{Comparison of occultation and infrared data (binaries):} An indirect way to detect the presence of satellites is comparing the sizes derived from infrared data and stellar occultations. Available thermal measurements for Centaurs are unresolved, providing only the total equivalent size. Among the infrared instruments used for thermal surveys of TNOs and Centaurs \citep[e.g. ][]{Muller2009}, \textit{Spitzer}/MIPS had an angular resolution between 6 and 40 arcsec \cite{Rieke2004}, equivalent to a spatial resolution of 22,000-145,000 km at 5 AU, while Herschel/PACS and SPIRE had 5 and 20-30 arcsec resolution, respectively \cite{Poglitsch2010, Griffin2010}. On the other hand, stellar occultations have inherently higher spatial resolution, down to sub-km level, allowing resolution of individual binary components. If the stellar occultation samples the main body of a multiple system, the multiplicity can be put in evidence as a smaller occultation size with respect to the thermal size. The presence of satellites due to the discrepancy in thermal and occultation sizes has been proposed in a few cases for TNOs, although there is no known case so far for Centaurs \cite{Rommel2023, Ortiz2020}.

\begin{figure}[b!]
\centering
\includegraphics[width=1\textwidth,clip,trim=0mm 0mm 0mm 0mm]{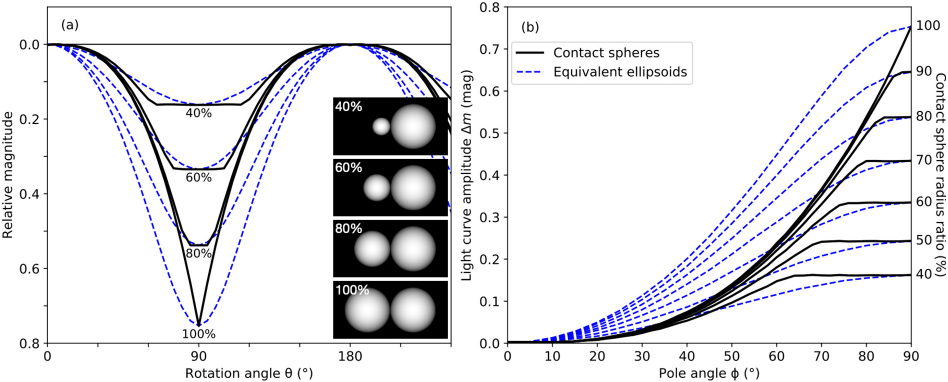}
\caption{(\textit{left}) Example model light curves for spherical (black) or ellipsoidal (blue) contact binaries of different relative sizes, exhibiting a characteristic V-shape. The pole angle is $90^{\circ}$. (\textit{right}) Models of peak-to-peak light-curve amplitude as a function of pole angle for contact binaries of different relative sizes. (Fig. 3 in \cite{Showalter2021})
\label{fig:lightcurves_1}}
\end{figure}

\begin{figure}[ht!]
\centering
\includegraphics[width=0.6\textwidth,clip,trim=0mm 0mm 0mm 2.5mm]{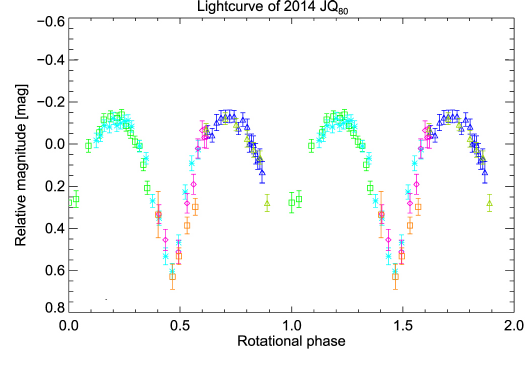}
\caption{Example of a rotational light curve for the contact binary candidate and Plutino 2014 JQ\textsubscript{80}, showing a characteristic V-shape. Data points of different colors were taken on different dates in May and June of 2017. (Fig. 3 in \cite[][]{Thirouin2018})
\label{fig:lightcurves_2}}
\end{figure}

\textit{Light curves (binaries):} The analysis of the shape, amplitude, and period of a rotational light curve is a complementary technique to search for satellites, by detecting multiple rotation periods, mutual events, or potential contact binaries. In the general case, a binary system can be revealed by the detection of the rotation period of a separate component, although this is rare in practice. Under favorable observation geometries, binaries can be detected by the occurrence of mutual events, while contact binaries can often be revealed by their characteristic V-shape light curve if the geometry is near edge-on (see Figs. \ref{fig:lightcurves_1} and \ref{fig:lightcurves_2}). The statistical analysis of light curves has been used to deduce a population of equal-size synchronous binaries among Jupiter Trojans \cite{Nesvorny2020} and to isolate a sample of contact-binary candidates among the TNO population, revealing an unexpectedly high fraction among Plutinos \cite{Thirouin2018}. An analysis of light curves to infer the shapes of bodies in the outer Solar System further found that the contact-binary fraction is not well constrained and could be very high \cite{Showalter2021}. 

So far, no contact binary has been reported among Centaurs or GPCs from light-curve analyses \cite{Thirouin2010}: the main limitation for this technique is the need for dense photometry of a somewhat faint population, which in turn requires many hours of scarce, large-size telescope time. Based on the cumulative distributions of light-curve amplitudes, Centaur light curves (adopting the Minor Planet Center definition) have been found to be statistically different from the rest of the TNOs, presenting an abundance of low-amplitude objects \cite{Showalter2021}. The origin of this difference is currently unknown.

\begin{figure}[t!]
\centering
\includegraphics[width=1\textwidth,clip,trim=0mm 0mm 0mm 0mm]{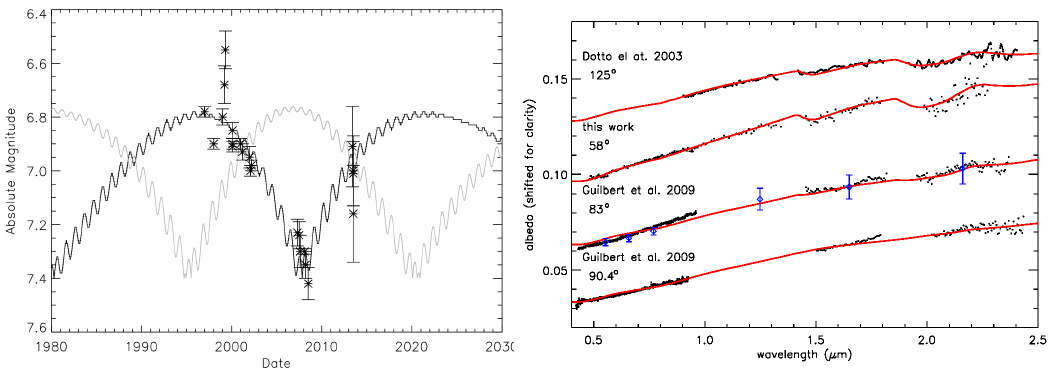}
\caption{Variability over time in Chariklo's brightness (\textit{left}) and reflectance spectra (\textit{right}). (\textit{left}) Model solutions for magnitude based on the two possible ring-plane poles are shown along with observational data: the darker line represents the preferred pole solution. (\textit{right}) Model spectra are shown with data taken at different epochs, demonstrating changes in the typical water-ice absorption features at 1.5 and 2 $\mu$m. These results were attributed to the changing aspect angle of Chariklo's water-ice ring particles. (Figs. 1 \& 5 in \cite{Duffard2014})
\label{fig:variability_1}}
\end{figure}

\textit{Magnitude and spectroscopic variations (rings or debris):} As the viewing geometry of substantial surrounding material changes with time, there should be corresponding absolute magnitude and spectral changes. For example, moving from edge-on to fully open rings, the brightness should increase and there should be a deeper water-ice feature in the spectra (assuming icy ring particles \cite[e.g. ][]{Hedman2013}). The existence of rings around Chariklo was bolstered, and those of Chiron proposed, based on such data  (see Fig. \ref{fig:variability_1} and \cite{Duffard2014, Ortiz2015}).

\section{Satellites}\label{sec:sats}
\subsection{Characteristics of known Centaur Binaries}

There are two known Centaur-like binaries: the GPCs Ceto-Phoycys and Typhon-Echidna. Characteristics of these systems are listed in Table \ref{tab:binary_char}. After initial discovery, additional observations by HST were able to determine their mutual orbits. This process involves making multiple observations with precise relative astrometry and then fitting an orbital model which includes the motion of the Earth (the HST-Earth relative position is unimportant), the heliocentric motion of the object, and the orbital parameters. Of particular interest is that the period and semi-major axis can be combined to measure the mass of the system; measuring the masses of individual components requires absolute astrometry which is generally not feasible. HST observations are the only practical way to measure astrometry for many binaries, so observations typically stop once a reasonably precise orbit can be inferred \citep[e.g.,][]{Grundy2007}. In some cases, there is a ``mirror degeneracy'' where the unknown direction of motion to/away from the observer leads to two precise orbits that cannot be distinguished in their orientation, though key parameters such as mass and eccentricity can be determined. Recently, it has become possible to employ more advanced dynamical and statistical methods to look for hints of ``non-Keplerian'' motion, where the orbits are not assumed to be fixed ellipses, but can include precession most typically induced by the oblateness of the components, known as $J_2$ \cite{Ragozzine2024,Proudfoot2024}. These analyses are still quite limited by the small amount of data. 

For Ceto-Phorcys, the binary period, semimajor axis, and eccentricity were determined even though the mirror degeneracy was not yet resolved because both the prograde and retrograde orbital solutions agreed \cite{Grundy2007}. The resulting total mass was combined with thermal modeling to return individual sizes for Ceto and Phorcys \cite{Grundy2007}. The system masses and sizes led to inferred densities, which did not include potential errors from systematic uncertainties in thermal modeling or the assumption of equal-albedo, equal-density spheres. The implied system albedo is somewhat intermediate between TNOs and JFCs, and the bulk density from \cite{Grundy2007} requires that there must be a rock component. On the other hand, the bulk density derived by \cite{SantosSanz2012} is significantly lower and the sizes larger. The inferred compositions depend significantly on the poorly-understood porosity.

\begin{deluxetable}{llcc}
\tablecaption{{Characteristics of Binary Giant Planet Crossers\textsuperscript{a}.\label{tab:binary_char}}}
\tablewidth{0pt}
\tablehead{
\colhead{Binary}&\colhead{Parameter}&\colhead{Value}&\colhead{Ref.}
}
\startdata
Ceto-Phorcys&&&\\
 &Mutual orbit period (days), $P$& $9.544\pm0.011$&\cite{Grundy2007}\\ 
 &Semimajor axis (km), $a$&       $1840\pm48$&\cite{Grundy2007}  \\ 
 &Eccentricity, $e$&              $0.008\pm0.008$&\cite{Grundy2007}\\
 &System Mass ($10^{18}$ kg), $M_{sys}$&    $5.5\pm0.3$&\cite{Grundy2007}\\ 
 & Bulk density (g cm\textsuperscript{-3})\textsuperscript{a}, $\rho$& $1.37^{+0.66}_{-0.32}$; $0.64^{+0.16}_{-0.13}$&\cite{Grundy2007}; \cite{SantosSanz2012}\\
 & System Albedo &               $0.084^{+0.021}_{-0.014}$; $0.056\pm0.006$&\cite{Grundy2007}; \cite{SantosSanz2012}\\
 & Radius of primary (km)\textsuperscript{b}  &    $87^{+8}_{-9}$; $223\pm10$&\cite{Grundy2007}; \cite{SantosSanz2012}\\
 & Radius of secondary (km)\textsuperscript{b} &    $66^{+6}_{-7}$; $171\pm10$&\cite{Grundy2007}; \cite{SantosSanz2012}\\
 Typhon-Echidna &&&\\
 &Mutual orbit period (days), $P$& $18.971\pm0.006$ & \cite{Grundy2008}\\
 &Semimajor axis (km), $a$&       $1628\pm29$  & \cite{Grundy2008}\\ 
 &Eccentricity, $e$&              $0.526\pm0.015$ & \cite{Grundy2008}\\
 &System Mass ($10^{18}$ kg), $M_{sys}$&    $0.95\pm0.05$& \cite{Grundy2008}\\
 & Bulk density (g cm\textsuperscript{-3})\textsuperscript{a}, $\rho$& $0.60^{+0.72}_{-0.29}$; $0.36_{-0.07}^{+0.08}$ &  \cite{Stansberry2012};\cite{SantosSanz2012}\\ 
 & System Albedo &               $0.06^{+0.041}_{-0.021}$; $0.044\pm0.003$ & \cite{Stansberry2012}; \cite{SantosSanz2012}\\
 & Radius of primary (km)\textsuperscript{b}  &    $137\pm30$; $162\pm{7}$ & \cite{Stansberry2012};\cite{SantosSanz2012}\\
 & Radius of secondary (km)\textsuperscript{b} &    $77\pm16$; $89\pm{6}$& \cite{Stansberry2012};\cite{SantosSanz2012}
\enddata
\tablecomments{\textsuperscript{a}Under the assumption of equal densities for each body. \textsuperscript{b}Assuming equal-albedo spheres.} 
\end{deluxetable}

Originally discovered by \cite{Noll2006}, Typhon-Echidna was then followed up for mutual-orbit determination by \cite{Grundy2008}. The system mass in Table \ref{tab:binary_char} assumes a nearly-Keplerian orbit. Although its size is similar to Ceto-Phorcys, the Typhon-Echidna binary has a larger eccentricity, which is very different from the effectively circular orbit of Ceto-Phorcys. Typhon's orientation was uniquely determined, and the system has a prograde orbit. The inferred system density is consistent with TNOs and JFCs of similar sizes.

Dynamical modeling of both binaries shows that there are not enough observations for a significant detection of non-Keplerian effects \cite{Proudfoot2024}. A single high-precision observation of Ceto would readily break the mirror degeneracy. Both binaries are good candidates for the discovery of slow non-Keplerian effects with only a small amount of new data, since any new observations would increase the baseline of observations by almost 20 years. The systems are also both on relatively compact orbits so that apsidal and nodal precessions due to non-spherical shapes could be significant and detectable. 


\subsection{Comparisons with Binary TNOs}

Centaur-like objects typically start as ``Hot" (Hot Classical, Resonant, and Detached) TNOs (see Chapter~3). Furthermore, approximately 30\% of Centaurs become JFCs \citep{Levison1997} and the transition from Centaur to JFC is enabled by a low-eccentricity dynamical Gateway \citep{Sarid2019}. Comparison of the binary frequency and characteristics between these populations is thus interesting to consider. However, it is very challenging to make direct comparisons of the binary fractions because there are observational detection biases as well as uncertainties in the sizes of these populations. 

Compared with the nearly 120 known binary TNOs, there are only two known binaries with Centaur-like orbits. Centaur binaries are not easy to detect. In 2006, direct imaging with HST detected Echidna around Typhon from a sample of eight objects, which included two Centaurs ((49036) Pelion and Bienor) and five other GPCs (29981, 33128, 54520, 60608, 87269) \cite{Noll2006}. A similar HST program targeting a sample of 12 Centaurs and six GPCs found Phorcys around Ceto \cite{Grundy2007}. \cite{Li2020} observed 23 Centaurs and 33 GPCs, finding no satellites while deriving an upper limit of 8\% binaries in their sample. It's worth noting that \cite{Li2020} included the known binary Typhon-Echidna in the sample, but they did not detect any satellites in those observations. The HST archive indicates imaging observations for 58 Centaurs and 109 GPCs, but the associated programs were focused on more than just searching for satellites. Due to the different instrument configurations, filters, and observation strategies, it is difficult to accurately determine the rate of binary systems within the Centaur and GPC populations. 

Ignoring biases and considering simply the frequency of binary detections versus numbers observed, the fractions of 0/58 Centaurs and 2/109 GPCs are interesting, especially given the estimate that $\sim$10\% of Hot TNOs are known to be binaries (and $\sim$30\% of Cold Classical TNOs) \cite{Noll2020}. There are also no known JFC binaries, although the majority of JFC nuclei are smaller than 10 km in effective diameter and have bilobate shapes \citep[e.g.][]{Thomas2009, Safrit2021}. The decreasing binary fraction from the Cold Classicals, to the more dynamically excited populations in the Trans-Neptunian region, to the Centaur and JFC populations, could very easily be a real effect and not due entirely to observational biases. 

Since the dynamical source of Centaur-like objects is known to have more observed binaries, we can speculate on the cause of the apparent differences in the binary fraction. One straightforward idea is that the binaries are affected by close encounters with giant planets, which are much more frequent for Centaur orbits compared to TNOs. Indeed, simulations have shown that widely-separated binaries are disrupted by scattering encounters with Neptune \citep{Parker2010,Nesvorny2019,Stone2021}. Most of these simulations have focused on the scattering that happens as Hot TNOs are emplaced in the early Solar System. That work is a helpful guide, but it is not as relevant as simulations that have looked at the present orbits of Centaurs specifically, such as \cite{Araujo2018}. 

In general, binary disruption will happen when external influences become more important than the mutual gravitational attraction. One metric for this is the observed semimajor axis of the satellite, $a_b$, relative to the mutual Hill sphere of the binary, $r_H$, which is the distance at which the gravitational influence of the Sun would overwhelm the binary's self-gravity. Values of $a_b/r_H$ of $\lesssim$1\% can be used to distinguish ``tight" from ``wide" binaries (although technically this metric is not strictly relevant for giant-planet encounters). Based on these dynamical simulations and considering that the known GPC binaries are both relatively tight systems, the general sense is that Ceto and Typhon have low probabilities of disruption after about 1 Myr, but their survival rate over tens of Myr (similar to their lifetimes) might be unlikely \citep[e.g.][]{Araujo2018}. Perhaps they are the fortunate survivors of an initial binary fraction that is consistent with the source population. 

Another possibility is that the comparison between Centaurs and TNOs is not appropriate because of observational biases. First, most discovery surveys find Centaur-like objects that are smaller than the corresponding TNOs simply due to the magnitude-limited nature of the observations. The binary frequency among TNOs as a function of size is not known, and perhaps the $\sim$100 km size range of Ceto and Typhon actually has fewer binaries than the sample as a whole. Furthermore, binary discovery is often limited by angular separation, so that separations of Centaur-like binaries would be barely resolvable at TNO distances in images (see Section \ref{subsubsec:directtechs}). For similar reasons, only binaries with larger $a_b/r_H$ can currently be detected in the TNO region using direct imaging. 

\begin{figure}[b!]
\centering
\includegraphics[width=1\textwidth,clip,trim=0mm 0mm 0mm 0mm]{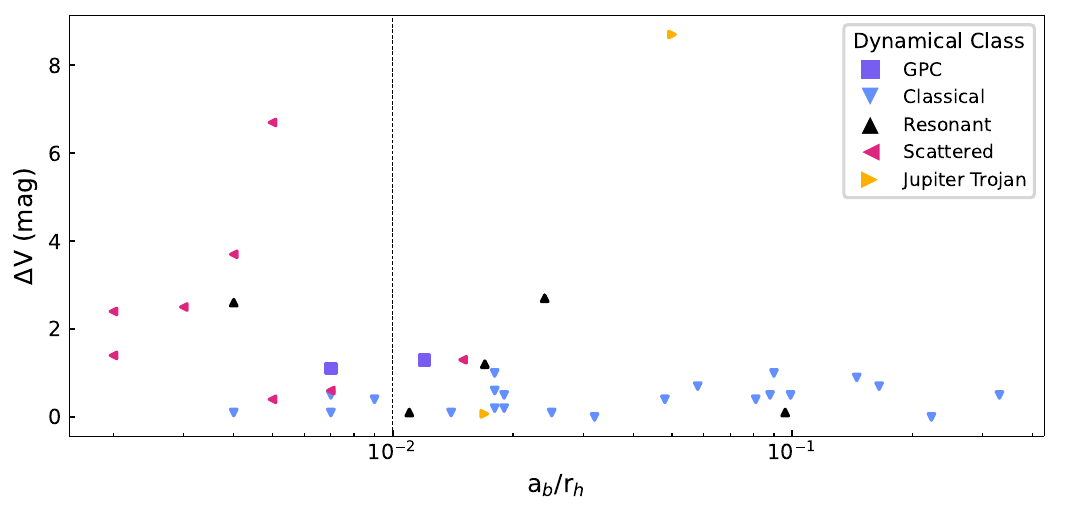}
\caption{Comparison of binary characteristics for Centaur-like objects (GPCs) with other dynamical classes in the outer Solar System. The horizontal axis is the ratio between the semimajor axis and the Hill radius and the vertical axis is the difference in apparent magnitudes of the binary components (as a proxy for their size difference). The vertical line indicates the $a_b/r_H$ = 1\% limit between ``tight" and ``wide" binaries. 
Data are from \cite{Grundy2019}.
\label{fig:binaries_2}}
\end{figure}

We can explore these possibilities by considering the properties of binaries from different dynamical classes. Gathering data on several binaries, we show in Fig. \ref{fig:binaries_2} the size ratio of the binary (estimated by the difference in magnitudes, $\Delta V$) versus the tightness of the binary measured by $a_b/r_H$, as a function of dynamical classification. 
The two GPC binaries have some of the smallest semimajor axes, appearing like an extension of tight TNO binaries. The GPC binary orbital characteristics are generally consistent with TNOs, including the moderate eccentricity of Typhon-Echidna. Though Centaur-like binaries are among the tightest known, there are a few TNO binaries that have been detected with similar characteristics, separations, and sizes, such as the Classical TNO (469514) 2003 QA\textsubscript{91}. There are thus observational biases against finding TNOs with orbits like the known GPC binaries, but they are not overwhelming. 

Nonetheless, relative observational bias implies that wide Centaur-like binaries must be more rare than tight Centaur-like binaries. This result seems consistent with estimates of dynamical destruction of these binaries. One potential observational signature that has not been explored is the discovery of Centaur or GPC ``pairs,"  separate objects on very similar orbits, which could only plausibly be formed by a disrupted binary. (We note that ultra-wide binaries that were the focus of \cite{Parker2010} are not particularly relevant to the Centaur-like population for two reasons. First, the widest binaries are generally Cold Classicals, which is likely a small source population for Centaurs. Second, there is new evidence that the observed ultra-wide binaries themselves could be modified from their original version by encounters with other TNOs \cite{Campbell2023}. The even more important influence of binary-binary encounters has not yet been studied in detail.)

When objects reach a point in their orbit that activity becomes significant, it is also worth considering whether non-gravitational effects on one or both components would be enough to influence the binary orbit. This is an open area for research. 
\subsection{Binary Candidates}
\label{subsec:otherbin}
Tidal evolution can slow the spins of binary components until they are synchronous. This is especially likely for the tightest binaries that would be difficult to resolve. As a result, long-period ($\sim$a few days) light curves may be indicative of an unknown binary. Two GPCs have been reported to have long-period light curves potentially indicative of binaries: (33128) 1998 BU\textsubscript{48} \cite{Sheppard2002} and 2010 WG\textsubscript{9} \cite{Rabinowitz2013}. Both have been observed with HST with no companion reported. The existence of confined rings (or the presence of a clear gap in a ring system) may also imply shepherd or other small moons, but these cannot be directly inferred from present data. 

\section{Rings}\label{sec:rings}

\subsection{Characteristics of proposed small-body ring systems}\label{subsec:ringchars}

Ring systems have been known around the giant planets for centuries: first Saturn's by Galileo Galilei in 1612 and, with the advent of space missions and large telescopes, later confirmed around Jupiter, Uranus, and Neptune \cite[e.g. ][]{Charnoz2018}. The data revealed different structures around the different planets, some of which is closely tied to satellites. Saturn has the most well-known system, surrounded by dense, tenuous, large, narrow, and arcing rings, with several moonlets in close connection with the ring particles. Notably, Saturn's narrow, dusty F-ring is located outside the Roche limit and its core is confined by the satellite Prometheus and precession \citep[e.g. ][]{Cuzzi2014}. Jupiter presents tenuous rings interior to the orbits of the Galilean satellites along with four small satellites, probably the source of these rings \citep[e.g. ][]{Ockert-Bell1999}. Uranus has several narrow rings, only a few-km wide, and the $\epsilon$~ring is confined by the shepherd satellites Cordelia and Ophelia \citep[e.g. ][]{Elliot1977,Goldreich1987,Porco1987}. Two decades later, two faint rings (the $\mu$ and $\nu$ rings) were discovered in HST images, lying outside the main ring system of Uranus \cite{Showalter2006}. Ring arcs were discovered at Neptune via stellar occultations \citep[e.g. ][]{Hubbard1986} and have continued to evolve \citep[e.g.][]{Souami2022}. Spacecraft images have shown that the arcs are in fact the densest parts of the Adams~ring, and they are likely radially confined by interactions with the moon Galetea \cite{Namouni2002} while small moonlets contribute to azimuthal confinement \cite{Renner2004,GiuliattiWinter2020}. In all of these systems, small satellites (or moonlets) play an important role in defining the structures of the rings. The gravitational interactions between the ring particles and a satellite create waves and gaps, and they can confine the edges of the rings or the arcs against disruptive forces, which tend to scatter the ring particles. 

\begin{figure}[b!]
\begin{center}
\includegraphics[width=0.5\textwidth]{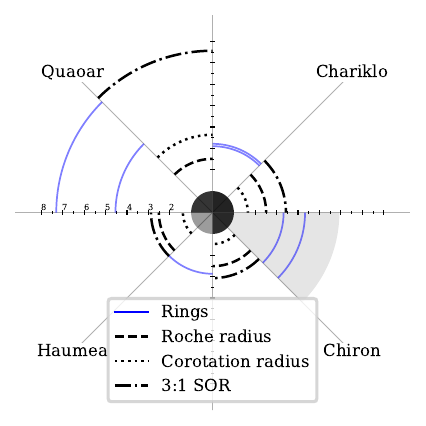}
\caption{\label{fig:ringcomp}Comparison of proposed small-body ring systems: (\textit{left}) TNOs Quaoar and Haumea, (\textit{right}) Centaurs Chariklo and Chiron. The radial distances are normalized by the volumetric-equivalent radii of the central bodies. The shading level of the central body represents the geometric albedo (from $\sim0.1-0.7$). The locations of rings or concentrations of material are indicated by solid, blue lines. The locations of the Roche radius, the 1:1 corotation radius, and the 3:1 spin-orbit resonance (SOR) are indicated with dashed, dotted, and dot-dashed lines, respectively (see Section \ref{subsec:roche} for the assumptions). For Chiron, the shaded area indicates the extension of a disk of material detected during a 2022 stellar occultation \cite{Ortiz2023}. Ring locations are from Table \ref{tab:rings} and the Roche radii were calculated using the equation for $a_{Roche}$ in Section~\ref{subsec:roche} adopting ${\rho_p}/{\rho_s}=2$ and  $\gamma=1.6$.}
\end{center}
\end{figure}

Four centuries after Saturn's rings were first observed, the discovery of two narrow, dense rings around the largest Centaur Chariklo from occultation data in 2013 sparked a new field of study: rings around small bodies \cite{Braga-Ribas2014}. Basic characteristics for Chariklo's rings, and those of the other known small-body systems, are given in Table~\ref{tab:rings} and plotted in Figure~\ref{fig:ringcomp}.  Chariklo's two-ring system has been well observed and characterized: stellar occultations between 2014-2020 (i) confirmed the existence of C1R and C2R, along with the circular ring solution and pole position from 2013, (ii) indicated that the inner ring varied azimuthally in width by up to 4.3 km while the outer ring varied in width up to 1 km, (iii) revealed W-shaped structure in the inner ring, and (iv) determined that Chariklo's shape is consistent with a triaxial ellipsoid with semi-axes $A=143.8^{+1.4}_{-1.5}, B=135.2^{+1.4}_{-2.8},C=99.1^{+5.4}_{-2.7} \, {\rm km}$ \cite{Berard2017,Leiva2017,Morgado2021}. The C1R ring-pole position has most recently been refined by 
\cite{Morgado2021}. Additionally, the lack of differences in 2017 visible ($0.45-0.65 \, {\rm \mu m}$) and red ($0.7-1.0 \, {\rm \mu m}$) occultation data at the 1-$\sigma$ level led to the conclusion that C1R contains particles mostly larger than a few microns in size \cite{Morgado2021}.

\begin{deluxetable}{rccccccl}
\rotate
\tablecaption{Characteristics of Proposed Small-Body Ring Systems.\textsuperscript{a} \label{tab:rings}}
\tablewidth{0pt}
\tablehead{
\colhead{Object} & \colhead{Rotation}  & \colhead{Location} & \colhead{Width}& \colhead{$e_r$\textsuperscript{c}} & \colhead{Ring Pole} & \colhead{Reference} \\
\colhead{} & \colhead{(hr)} &\colhead{(km)\textsuperscript{b}} & \colhead{(km)} & \colhead{} & \colhead{$\alpha_p$, $\delta_p$ (deg, J2000)\textsuperscript{d}} & \colhead{}
}
\startdata
(2060)~Chiron\textsuperscript{e} & 5.92& $325\pm16 \, {\rm km}$ & - & 0  &$160.2_{-12.4}^{+13.2}$, $27.8_{-13.2}^{+13.9}$&\cite{Marcialis1993,Ortiz2023} \\
 && $423\pm11 \, {\rm km}$ & - & 0 & &\\
(10199)~Chariklo C1R &7.004& $385.9\pm0.4$ & $\sim$6.9\textsuperscript{f}&$0.005<e_r<0.022$& $151.03\pm0.14$, $41.81\pm0.07$ &\cite{Fornasier2014,Morgado2021,Melita2017} \\
 C2R && $399.8\pm0.6$ & $\sim$3.5\textsuperscript{g} & \textless0.017 & $150.91\pm0.22$, $41.60\pm0.12$& \cite{Braga-Ribas2014, Morgado2021}\\
\hline
(50000)~Quaoar&& && & &\\
Q1R (epoch 2018-2021) &8.84; 17.69& $4148.4\pm7.4$ &8; 20-340\textsuperscript{h} & 0 & $259.82\pm0.23$, $53.45\pm0.30$ & \cite{Ortiz2003,Morgado2023} \\
Q1R (epoch 2022)&& $4,057.2\pm5.8$ &5; 80-100\textsuperscript{i} & 0&  &\cite{Pereira2023,Pereira2024} \\
Q2R (epoch 2022)&& $2, 520\pm20$ & $\sim$10\textsuperscript{j}& 0  &&\cite{Pereira2023,Pereira2024}\\
(136108)~Haumea&3.9& $2287^{+75}_{-45}$ & $\sim$70\textsuperscript{k} & 0  & $285.1\pm0.5$, $-10.6\pm1.2$ &  \cite{Rabinowitz2006,Ortiz2017} 
\enddata
\tablecomments{
\textsuperscript{a}{Centaurs are above the solid line and TNOs below. Opacities and optical depths are given in Table \ref{tab:ringcomps} and discussed in Section \ref{subsec:tau}.}
\textsuperscript{b}{Radial distance.}
\textsuperscript{c}{Ring eccentricity, $e_r$: values of 0 are for rings assumed to be circular.}
\textsuperscript{d}{Right ascension and declination of the ring-plane pole, $\alpha_p$ and $\delta_p$, for the preferred pole solutions from \cite{Ortiz2023,Morgado2021,Pereira2023,Ortiz2017}.}
\textsuperscript{e}{Locations of concentrations of material, within an underlying $\sim 580 \, {\rm km}$ disk, from the most recent occultation observation in 2022. Data from previous occultations showed, among other detections, two distinct features at $\sim300$ and $\sim310$ km with widths ranging between 2.6 and 4.4~km (in 2011 \citep{Sickafoose2020}) and between $\sim344-364 \, {\rm km}$ with widths ranging from 2.2-4.5~km (in 2018 \citep{Sickafoose2023}).}
\textsuperscript{f}{The mean width for C1R was reported as 6.9 km in the text (not the abstract), ranging between 4.8 and 9.1~km \cite{Morgado2021}.} 
\textsuperscript{g}{Measured widths of C2R were $3.6^{+1.3}_{-2.0} \, {\rm km}$ and $3.4^{+1.1}_{-1.4} \, {\rm km}$ in \cite{Braga-Ribas2014}. A typical radial width was not given for more recent observations, but measured values ranged from $0.095^{+0.015}_{-0.010} \, {\rm km}$ to $3.72^{+0.63}_{-0.53} \, {\rm km}$ \cite{Morgado2021}.}
\textsuperscript{h}{When detected, the adopted radial width was 8~km for the ``densest part of the ring," while other measured widths ranged from $21.34\pm3.13$ to $336.34\pm 23.81 \, {\rm km}$ \cite{Morgado2023,Morgado2024}.}
\textsuperscript{i}{When detected, the ``dense part" of the ring was consistent with a Lorenztian shape extending 60~km with full-width-at-half-maximum of 5~km and, from the best light curves, the ``tenuous part" was typically 80-100~km \cite{Pereira2023,Pereira2024}.}
\textsuperscript{j}{Assuming coplanar with Q1R. The typical width was given as 10 km, with measured widths ranging from $6.8\pm0.8$ to $16.1\pm3.3 \, {\rm km}$ \cite{Pereira2023,Pereira2024}.}
\textsuperscript{k}{Adopted width of 70~km, with measured ingress and egress values of $\sim 74\, \& \, 44 \, {\rm km}$ from the one site at which the ring profile was resolved \cite{Ortiz2017}.}
}
\vspace{-30pt}
\end{deluxetable}

The situation at Chiron is not quite as straightforward. After its discovery in 1977, Chiron presented cometary-like activity \citep[e.g. ][]{Meech1989,Luu1990,Elliot1995}. Data from a two-chord stellar occultation in 2011 were originally interpreted as showing a shell of material \cite{Ruprecht2015}: by combining these results with long-term photometry and spectroscopy, a two-ring system at $\sim324 \, {\rm km}$ similar to that at Chariklo was proposed \citep{Ortiz2015}. Stellar-occultation data in 2018 and 2019 allowed better constraint of the size and shape of Chiron ($A=126\pm{22},B=109\pm{19},C=68\pm{13} \, {\rm km}$, assuming a Jacobi equilibrium shape and considering the light-curve amplitude \cite{Braga-Ribas2023}). Analysis of stellar occultations in 2018 and 2022 suggested that the proposed rings are rather surrounding material that is evolving in a very short period of time \cite{Ortiz2023,Sickafoose2023}. Based on the most recent observations, \cite{Ortiz2023} proposed that Chiron possesses a tenuous disk of material roughly 580~km wide, with concentrations of material located at $325 \pm{16} \, {\rm km}$ and $423 \pm{11} \, {\rm km}$ and pole ecliptic coordinates of $\lambda=151^\circ \pm8^\circ$ and $\beta=18^\circ \pm11^\circ$.

From a stellar occultation in 2017, a 70-km wide ring was also discovered around the TNO Haumea (Table~\ref{tab:rings} and \citep{Ortiz2017}). Compared to the Centaurs, Haumea is more distant ($q=35$~AU and $a_\odot=43$~AU), is an order of magnitude larger, and has a highly elongated triaxial shape with semi-axes $A = 1161 \pm 30$, $B = 852 \pm 4$, and $C = 513 \pm 16$~km \cite{Ortiz2017}. Haumea's ring is coplanar to the equator and to the orbit of the outer satellite Hi'iaka. The two known satellites are located tens of thousands of kms away from Haumea \citep[e.g.][]{Brown2007}; therefore, their effects are minimal and the non-spherical shape of Haumea likely plays an important role in the dynamics of the particles (see Section~\ref{subsec:ringmodels}).

Most recently, a ring system was reported around the TNO Quaoar (Table~\ref{tab:rings} and \cite{Morgado2023}). Quaoar's light curve has been interpreted to be single- or double-peaked, yielding a rotation rate of 8.84 or 17.69~hr \cite{Ortiz2003}. A non-homogeneous ring Q1R, with more non-detections in the occultation data than detections, presents an interesting scenario. It is located significantly beyond the Roche limit, regardless of the assumptions (see Section \ref{subsec:roche}). 
A second proposed ring, Q2R interior to Q1R, is likewise beyond the Roche limit \cite{Pereira2023}. The variable nature of the occultation detections around Quaoar and the highly unusual locations of reported surrounding material make this a very compelling target for future study.

The locations of the small-body ring systems are intriguing because outside the Roche limit, particles can accrete into satellites in a very short period of time (additional discussion in Section~\ref{subsec:roche}). Furthermore, the discovery of thin rings around small-bodies is surprising because over relatively short timescales, ring material will naturally disperse ($<1 \, {\rm Myr}$, e.g. \citep{Braga-Ribas2014,Pan2016}). Understanding is thus required of both how the rings can survive far from the Roche limit and how thin rings can be confined over long periods of time. Proposed mechanisms for formation and confinement are discussed in Section \ref{subsection:ringformation}.

In terms of composition, small-body rings are thought to be made of water ice mixed with other materials (as are Saturn's rings \cite[e.g.][]{Nicholson2008}). For Chariklo and Chiron, correlations between proposed ring opening angles, visual brightness, and water-ice spectral features have implied that the rings contain water ice (see e.g. Fig. \ref{fig:variability_1} and \cite{Duffard2014, Ortiz2015}): a best-fit model for Chariklo's rings was 20\% water ice, 40-70\% silicates and 10-30\% tholins, with small amounts of amorphous carbon \cite{Duffard2014}. Mixtures of 20\% or 30\% water ice and 80\% or  70\% refractory material are compatible with near-infrared spectra for Chariklo and Chiron, respectively \cite{Groussin2004}, the water ice of which might all be in rings or surrounding material. In addition, observations of Quaoar's rings are consistent with particles that have icy density (e.g. that of small inner saturnian satellites), and the rings have been modeled as frost-covered ice \cite{Morgado2023}. 

There are four Centaurs for which upper limits have been placed on putative rings: Echeclus, SW1, Bienor, and 95626. Three stellar occultations by Echeclus have been published, with the best light curves at each epoch having $3\sigma$ upper limits on observed optical depth of 0.035 for features of width 2.6~km in 2019, 0.075 for features of width 2.9~km in 2020, and 0.08 for features of width 1.2~km 2021 \cite{Pereira2024b}. For 100-km wide features, the upper limits for all observations ranged from 0.005 to 0.045 \cite{Pereira2024b}. Initial results from three successful occultation observations in 2022 and 2023 by SW1 did not show any obvious signs of extended structures \cite{Buie2023}. Stellar occultation data for Bienor in 2019 returned the strongest constraint that a ring of width $<14.1 \, {\rm km}$ and opacity of 50\% would not be detected \citep{Fernandez-Valenzuela2023}. Finally, a stellar occultation by 95626 in 2017 returned the $3\sigma$ ring constraint (within the positive observations) that widths $>3.7$ and $>1.8$~km could have been detected for opacities of 50\% and 100\%, respectively \cite{Santos-Sanz2021}. For the best data near the limb, rings $>1.1$ or $>0.5 \, {\rm km}$ could have been detected for opacities of 50\% and 100\%, respectively \cite{Santos-Sanz2021}. These results rule out Chariklo-like rings at Echeclus and likely SW1, but the possibility of narrow and thin ring systems was not discarded for Bienor or 95626 \cite{Pereira2024b,Fernandez-Valenzuela2023,Santos-Sanz2021}.

\subsection{Published Values of Ring Optical Depths}\label{subsec:tau}
It is worth noting that observed values of optical depths for Centaur rings have not been consistently reported in the literature. A stellar occultation light curve provides a measurement of the transmission, $T$, or the fractional amount of light observed when a star is blocked by ring material. Ring opacity along the line of sight, $p$, is related to transmission through $p=1-T$. Transmission is related to optical depth along the line of sight, $\tau_0$, through $T=e^{-\tau_0}$ when assuming the ring is a gray screen many particles thick \citep[e.g., ][]{Elliot1984}. To convert to normal optical depth, $\tau_{N}$, which provides the most physically-relevant information about the ring, requires knowing the ring opening angle, $B$, where $\tau_{N}=\tau_{0} sin\vert B \vert$ for a polylayer ring \citep[e.g.][]{Elliot1984}. 

For Haumea, only line-of-sight opacities were reported \citep{Ortiz2017}. For Chiron, published values of optical depths for surrounding material were based on this standard calculation of transmitted flux \citep{Sickafoose2020}. In contrast, reported optical depths for Chariklo's rings and the proposed ring at Quaoar have included a factor of two. The relationship between transmission and optical depth for observations of Chariklo was defined as $T=e^{-2\tau_0}$ \citep{Braga-Ribas2014, Berard2017}. For Quaoar, the normal optical depth was defined as $\tau_{N}=sin\vert B \vert\ \tau_0$/$ 2$ \citep{Morgado2023, Pereira2023} (noting that an additional factor of two was applied in the original manuscripts but revised in corrigenda \cite{Morgado2024,Pereira2024}).  According to \citep{Berard2017,Morgado2024}, the factor of two stems from \citep{Cuzzi1985}, who found that the fill factor, $\tau_N$/$Q_{e}$, had been underestimated for the Uranian rings and that the likely value of the Mie coefficient for those observations was $Q_{e}=2$. Consistently, diffraction for a ``zebra-striped" screen ($\tau=0$ or $\infty$ for alternate stripes) returns an optical depth that is twice that of the optical depth produced by a gray screen of equivalent width (as described in Section V.A of \cite{French1986}).

The Mie coefficient comes into play when using optical depth to derive physical properties, such as particle sizes, densities, or reflectivities \citep[following, e.g., Eq. 1 in][]{Cuzzi1985}. The ring characteristics of the small bodies are not well known; therefore, Table \ref{tab:ringcomps} contains ring opacities and optical depths calculated using the standard, gray-screen definition for all four systems so that nominal values can be compared. There is a wide range of optical depth measurements, especially when including error bars. We emphasize that the diffraction effects do need to be taken into consideration to infer characteristics of ring particles from the observed optical depths \citep{French1986}: if the rings are more ``zebra-striped" than a polylayer gray screen, the factor of two described above comes into play and the \textit{effective} optical depths can reach half of those in Table \ref{tab:ringcomps}. Note that Table \ref{tab:ringcomps} does not contain the most recent stellar occultation by Chiron in 2022, as neither opacities nor optical depths were reported \cite{Ortiz2023}.

\begin{deluxetable}{llccccccc}
\rotate
\tablecaption{Consistently-Derived Opacities and Optical Depths for Proposed Small-Body Rings\tablenotemark{a}.\label{tab:ringcomps}}
\tablewidth{0pt}
\tablehead{
\colhead{Body}& \colhead{Observation} &\colhead{Line-of-sight} & \colhead{Line-of-sight}& \colhead{Ring Opening} & \colhead{Range of Normal}& \colhead{Nominal Normal}& \colhead{Equivalent}\\
\colhead{}& \colhead{Date}& \colhead{Opacity} & \colhead{Optical Depth}& \colhead{Angle} & \colhead{Optical Depths}& \colhead{Optical Depth} & \colhead{Width}\\
\colhead{}& \colhead{(UT)}& \colhead{$(p)\tablenotemark{b}$} & \colhead{($\tau_0$)\tablenotemark{b}}& \colhead{($B$;$^\circ$)\tablenotemark{c}}&\colhead{$(\tau_{N})\tablenotemark{b}$} &\colhead{$(\tau_{N})\tablenotemark{d}$}&\colhead{($E_{W}$; km)\tablenotemark{e}}
}

\startdata
Chariklo C1R\textsuperscript{f}& 2013 Jun 03 & $0.67-0.81$ & $1.11-1.65$ & 33.77& $0.62-0.92$& 0.8& $2.3-3.3$\\
Chariklo C2R\textsuperscript{f}& 2013 Jun 03 & $0.13-0.35$ &$0.14-0.43$ & 33.77 & $0.08-0.24$& 0.12& $0.1-1.0$\\
Chiron F1\textsuperscript{g}& 1994 Mar 09 & $0.57-0.63$ & $0.85-0.99$ & –57.9 & $0.72-0.84$& 0.78&$2.7-4.9$\\
Chiron inner\textsuperscript{h}& 2011 Nov 29 & $0.41-0.72$ & $0.52-1.28$ & 59.6 & $0.45-1.10$& 0.73&$0.8-3.3$\\
Chiron outer\textsuperscript{h}& 2011 Nov 29 & $0.33-0.62$ & $0.40-0.97$ & 59.6 & $0.35-0.84$& 0.63&$0.8-2.2$\\
Chiron\textsuperscript{i}& 2018 Nov 28&$0.10-0.43$ & $0.11-0.57$ & 47.3 & $0.08-0.42$& 0.25& $0.2-1.6$\\
Haumea\textsuperscript{j}& 2017 Jan 21& $0.55-0.56$ & $0.80-0.82$ & –13.8 & $0.190-0.196$& 0.193& $5.9-9.7$\\
Quaoar Q1R\textsuperscript{k}&2019 June 05\textsuperscript{1} & $0.04-0.11$ & $0.04-0.12$ & –19.5 & $0.012-0.039$& 0.025& $1.0-7.8$\textsuperscript{m}\\
&2021 Aug 27\textsuperscript{n}& $0.66-1.00$ & $1.09-16.10$ & –19.3 & $0.36-5.32$& 2.84& $1.1-3.9$\textsuperscript{p}\\
& 2022 Aug 09\textsuperscript{q} & $0.03-0.85$& $0.03-1.87$ & –20.0 & $0.01-0.64$& 0.22 &$0.4-7.7$\textsuperscript{r}\\
Quaoar Q2R& 2022 Aug 09\textsuperscript{q} & $0.02-0.06$ & $0.02-0.06$& –20.0 & $0.007-0.022$& 0.015& $0.03-0.11$\textsuperscript{r}\\
\enddata
\tablecomments{\textsuperscript{a}{Assuming fractional transmission $T=e^{-\tau_0}$, $p=1-T$, and $\tau_{N}=\tau_0 sin \vert B \vert$ for a polylayer ring, following \cite{Elliot1984}. }
\textsuperscript{b}{Full range of values when detected, including published error bars. }
\textsuperscript{c}{Based the preferred ring-pole solutions from \cite{Braga-Ribas2014,Ortiz2015,Ortiz2017,Morgado2023} and published coordinates at the time of the occultations.}
\textsuperscript{d}{Nominal value, either as cited or the average of the measured values. }
\textsuperscript{e}{Defined as $E_W=W p_N$, where $p_N=p sin\vert B \vert$ and the minimum and maximum radial ring widths, $W$, are from published values including errors.}
\textsuperscript{f}{For each of the two rings, recalculated based on the values of $\tau_{N}$ in \citep{Braga-Ribas2014}. }
\textsuperscript{g}{Assuming minimum and maximum line-of-sight optical depths for the strongest feature (F1) resolved in the ``KAO, optical" data in \citep{Elliot1995}. 
}
\textsuperscript{h}{Assuming minimum and maximum line-of-sight optical depths for each of the two features in the full-resolution ``FTN" data from \citep{Sickafoose2020}. }
\textsuperscript{i}{Assuming minimum and maximum line-of-sight optical depths for all of the detected features from \citep{Sickafoose2023}.}
\textsuperscript{j}{Assuming line-of-sight ring opacities from the resolved ring profiles in \citep{Ortiz2017}.}
\textsuperscript{k}{Characteristics from the events with multiple detections are listed here: two additional events were reported in \citep{Morgado2023} .}
\textsuperscript{l}{Using the minimum and maximum apparent opacities on this date, specifically for ``GTC $i_s$ (bef.)" and ``GTC $r_s$ (aft.)" in the corrected version of Extended Data Table 2 \citep{Morgado2023,Morgado2024}.}
\textsuperscript{m}{Using the minimum and maximum equivalent widths on this date, specifically for ``GTC $r_s$ (aft.)" and ``GTC $r_s$ (bef.)" in \citep{Morgado2023,Morgado2024}.}
\textsuperscript{n}{Using the minimum apparent opacity on this date for ``S. Valley (bef.)" and the maximum apparent optical depth from ``Reedy Creek (bef.)" (we use the optical depth because the maximum apparent opacity for this date is given as $1.000+0.000$, which returns infinite optical depth$^a$), in \citep{Morgado2023,Morgado2024}. No site observed ring material on egress.} 
\textsuperscript{p}{Using the minimum and maximum apparent opacities and ring widths for ``Reedy Creek (bef.)" in \citep{Morgado2023,Morgado2024}.}
\textsuperscript{q}{Using the minimum and maximum $\tau_{N}$ values for ``Gemini (z')" for Q1R and ``CFHT" ingress and ``Gemini z'" egress for Q2R in \citep{Pereira2024} (line-of-sight values were not provided). Note that Q2R was not detected during ingress for ``Gemini (r')".}
\textsuperscript{r}{Using the minimum and maximum $W p_N$ from ``Gemini (z')" on ingress and "TUHO" on egress for Q1R, and from ``CFHT" on ingress and "Gemini (r')" on egress for Q2R. Note that fitted equivalent width values in \citep{Pereira2024} differ from the simple equation used here.}
}
\vspace{-30pt}
\end{deluxetable}


\subsection{Consideration of the Roche Limit}\label{subsec:roche}

By most classical calculations, the rings around small-bodies occur near or outside their Roche limits: the Quaoar system is exceptionally far, at more than twice this distance (see Fig. \ref{fig:ringcomp} and \cite{Morgado2023}). The Roche limit is the distance from the central body inside which tidal forces prevent the accretion of ring particles or can disrupt an orbiting body. It is expected that a ring system can exist inside this limit and a cluster of particles can accrete into a small moon beyond it. However, this limit depends on the mass of the body as well as the parameters of the object to be disrupted, such as density and internal material strength \cite{Tiscareno2013}.

For triaxial bodies, the Roche radius can be defined as $a_{Roche}=(4 \pi ABC\frac{\rho_p}{\gamma \rho_s})^{1/3}$, where $A$, $B$, and $C$ are the semimajor axes of the primary, $\rho_p$ is the density of the primary, $\rho_s$ is the density of the secondary, and $\gamma$ is a dimensionless geometrical parameter describing the sphericity of the secondary \cite{Tiscareno2013}. Values of $\gamma$ typically range from 0.85 to 1.6, the former as a limiting value for the equilibrium shape of an incompressible fluid (that may not be achieveable for solid materials) and the latter representing a fully-filled Roche lobe with uniform density \citep[e.g. ][]{Tiscareno2013, Porco2007}. From this equation, for example, in order for the Roche limit of Chariklo to be beyond the location of the rings, the density of the orbiting material needs to be exceptionally low with respect to the primary and/or $\gamma$ needs to be low \cite{Melita2017}. The uncertainties in Chariklo's density and mass lead to a fairly wide range of possible Roche limits. Given the mass estimate for Chariklo of $6-8\times10^{18} \, {\rm kg}$ \cite{Leiva2017}, assuming the ring material has density $\rho_s=400 \, {\rm kg\,m^3}$ (the value adopted for the small inner satellites of Saturn), and substituting the primary's mass $M_p=(4\pi/3)(ABC)^3\rho_p$, $a_{Roche}$ at Chariklo ranges between $304-413 \, {\rm km}$.

As discussed in \cite{Tiscareno2013}, for a ring system it is better to consider the critical density ($\rho_{Roche}$) instead of the critical distance, which is given by $\rho_{Roche} =3M_p$/$\gamma a_b^3$, where $a_b$ is the semimajor axis of the secondary orbit \citep{Porco2007}.  For a given value of $a_b$, this is the critical density at which the object's size fills its region of gravitational dominance. As an example, for Quaoar Q1R, the  value of  $\rho_{Roche}\approx30 \, {\rm kg\, m^{-3}}$, assuming $\gamma = 1.6$ \cite{Morgado2023}. However, this value  corresponds to very porous or fluffy material. Instead, the classical value of the Roche limit at Quaoar was found to be $\sim$1780~km assuming ring-particle density of $400 \, {\rm kg\, m^{-3}}$ \cite{Morgado2023}. \cite{Morgado2023} suggested that elastic collisions can maintain such a ring beyond the Roche limit. Material beyond the Roche limit can also be prevented from accreting if its radial velocity dispersion increases due to external perturbations. A spin-orbit resonance or shepherd satellite(s) might play roles in maintaining the observed rings \cite{Sicardy2019,Sickafoose2024}. At Quaoar, the 6:1 mean-motion resonance with Weywot may also be involved \cite{Morgado2023, Pereira2023}.

\subsection{Ring Modeling}\label{subsec:ringmodels}
Thin rings around giant planets or small bodies do not survive throughout the age of our Solar System \citep[e.g. ][]{Tiscareno2013,Pan2016}. These rings require confinement mechanisms to prevent the spreading and/or a source to replenish the lost particles. Resonances between particles and the central body or nearby satellites can maintain ring confinement for a longer time. However, Neptune's arcs have been found to be transient \citep[e.g. ][]{DePater2018}. There have been theoretical and numerical studies of the evolutionary dynamics in the small-body systems, more recently including the gravitational effects on ring particles around prolate, small, central bodies. 

Since all the large-body ring systems are close to the giant planets (which are oblate bodies), the gravitational effect on the ring particles can be modelled using the $J_n$ terms. It is well-known that the oblateness of the central body provokes large short-term variations on the osculating orbital elements of a particle around the primary, which can be corrected through the use of the geometrical elements \cite[e.g. ][]{Renner2006}. However, for ellipsoid bodies, such as Chariklo and Haumea, in addition to the $J_n$ coefficients, the $C_{22}$ term (ellipticity of the equatorial region of the primary) has to be added in the gravitational potential. The $C_{22}$ gravity coefficient can induce an increase in the eccentricity of the particle, which can be corrected using an appropriate choice of initial conditions \cite{Ribeiro2021}. Unfortunately, only adding terms in the geometrical-elements algorithm does not solve the problem. To deal with this, \cite{Ribeiro2021} developed a set of empirical equations as a function of $C_{22}$ which ensures that the particle will perform the nominal orbit around a prolate body. 

Narrow, eccentric rings like those seen at Uranus have apse alignment due to self-gravity \cite{Goldreich1979}: a simple model that combined the ellipticity of the central body and the particles' self-gravity to maintain apse alignment was proposed for Chariklo \cite{Pan2016}. This work determined a mass for the inner ring  $10^{16} \, {\rm g}$, a typical particle size of a few meters, and a spreading time of $10^5 \, {\rm yr}$. A theoretical model for apse-alignment was also developed to constrain Chariklo’s ring surface density and eccentricity gradient, as well as the relative, minimum mass and location for a putative satellite \cite{Melita2020}. It turns out that the rings around Chariklo and Haumea are close to the 3:1 spin-orbit resonances (SORs). \footnote{Here, we define SOR literally, as spin:orbit or \textit{number of rotations}:\textit{number of orbital revolutions}, while noting that some works reverse these values (effectively providing orbit:spin resonance numbers).} 
These locations prompted studies of the effects of resonances between the spin of a non-axisymmetric body (elongated or with a topographic feature) and the orbital motion of the particles \cite{Sicardy2019, Sicardy2020}. The shape of the central body pushed the ring material beyond the 2:1 SOR, clearing the region nearby: particles located inside the corotation radius migrated towards the central body, while those located outside were pushed outside the 2:1 resonance. Thus fast rotators, where the 2:1 SOR is within the Roche limit, could be most likely to host rings \cite{Sicardy2019}.

\begin{figure}[t!]
\centering
\includegraphics[width=0.48\textwidth,clip,trim=0mm 0mm 0mm 18mm]{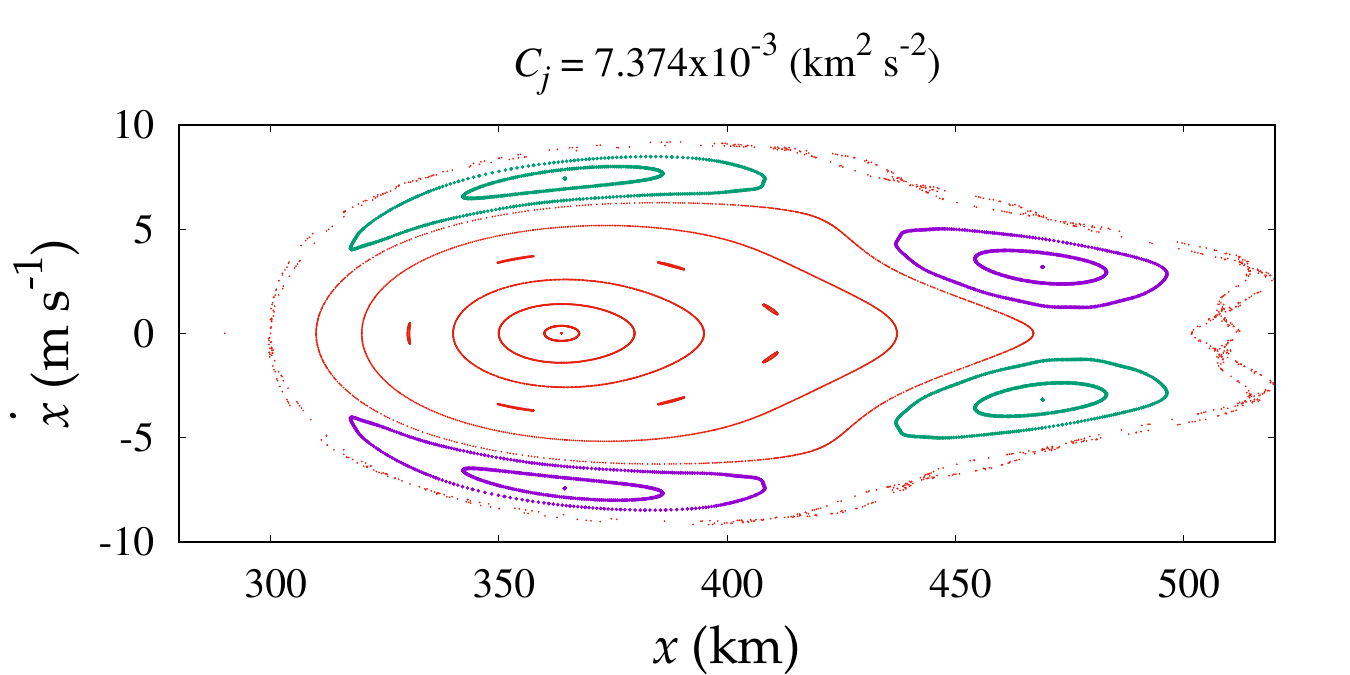}
\includegraphics[width=0.48\textwidth,clip,trim=0mm 0mm 0mm 18mm]{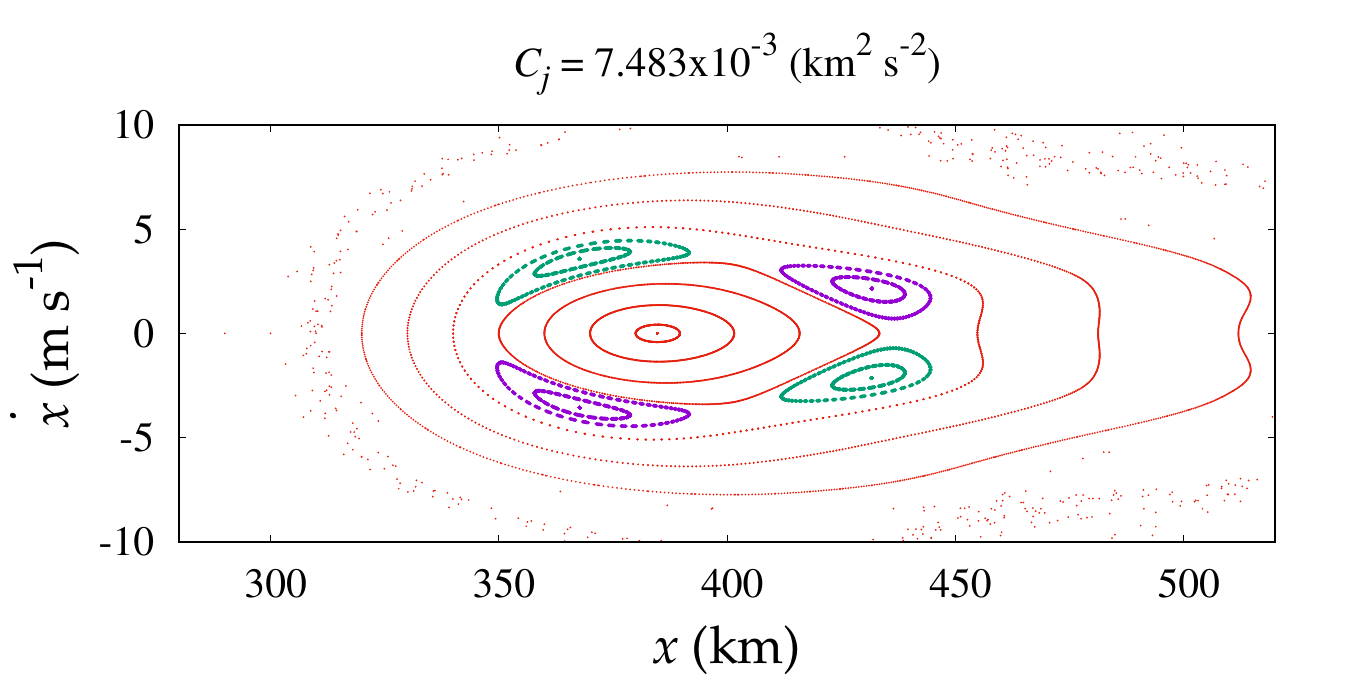}
\caption{Examples of  Poincaré Surface of Sections (PSSs) within Chariklo's 3:1 SOR. A family of first kind periodic orbits is responsible for the quasi-periodic orbits shown as red islands. Purple and green represent two families of 3:1 resonant periodic orbits (periodic orbits of the second kind). The Jacobi constants for these plots were (\textit{left}) $C_j=7.374\times10^{-3} \,{\rm km^2s^{-2}}$ and (\textit{right}) $C_j=7.483\times10^{-3} \,{\rm km^2s^{-2}}$.
Figure  extracted from Fig.~2 in \cite{GiuliattiWinter2023}.
\label{fig:PSS}}
\end{figure}

Several studies have analyzed the stability of the surrounding region around prolate bodies through the powerful technique of the Poincar\'e Surface of Section (PSS), which allows identification of stable and chaotic regions as well as the locations of resonances. Figure~\ref{fig:PSS} shows two PSSs for the Chariklo system covering the 3:1 SOR for different values of the Jacobi Constant. In these plots, resonant orbits are represented by fixed points, quasi-periodic orbits are represented by islands around the fixed points, and the points spread over the section are identified as chaotic trajectories. From a set of PSSs, a semimajor axis versus eccentricity map can be generated that identifies the locations and widths of the stable regions and resonances \citep[e.g.][]{Winter1997,Ribeiro2021,GiuliattiWinter2023}.

Application of the PSS technique suggested that Haumea's ring is located in a stable region associated with first-kind periodic orbits, and not with the 3:1 SOR \cite{Winter2019}. Similar results were found for Chariklo's C1R, which is located in a stable region due to first-kind period orbits, while C2R is located in an unstable region \cite{GiuliattiWinter2023}. Only large values of the eccentricity of C2R, and consequently a larger width, locate C2R in a stable region, prompting consideration of a three-moon system to confine the rings \cite{GiuliattiWinter2023}. 

Additional analyses for a body with a topographical feature a few kilometers long (a mass anomaly like that proposed by \cite{Sicardy2020} for Chariklo) used the adapted pendulum model and PSS to verify that the resonance locations are mainly affected by the mass and the spin-period of the primary, while the topographical feature can influence the width of the resonance \cite{Madeira2022}. For the Chariklo system, the results suggested that the stability of the C1R is associated with the periodic/quasi-periodic first kind orbits. On the other hand, an investigation of various spin rates and ellipticities of central bodies for Haumea, Chariklo, and five hypothetical systems found that periodic first-kind orbits are present in all systems with almost zero eccentricity of the particles and that resonant orbits have high eccentricities \cite{Ribeiro2023}. 

Finally, global N-body simulations have been carried out to model the Chariklo system, with the conclusions that Chariklo should be more dense than the ring material and that the existence of narrow rings implied smaller than m-sized particles or the existence of shepherd satellites \cite{Michikoshi2017}. The most recently published N-body studies, developed from existing models for Saturn's rings, find that a single shepherd moon is capable of confining material into the thin widths and locations observed at Chariklo \cite{Sickafoose2024}. Such a moon would only need to be a few km in diameter, below current imaging-detection limits.

\section{Debris}\label{sec:debris}
Any active Centaur would be likely to have some level of debris environment from dust to boulders depending on the amount of activity and how recently an outburst may have occurred. Furthermore, given the mass of some of these objects, the larger Centaurs may have long-lived orbital debris leftover from periods of activity \cite{Fink2021}. Debris can be detected and characterized through a variety of techniques, as also discussed in Chapters 8 and 10. The large-scale debris considered in this chapter could be detected through stellar occultations, if it has sufficiently high optical density or large orbiting bodies. So far, observations of occultations by active Centaurs have only detected significant debris in Chiron's near environment. 

As noted in section \ref{subsec:ringchars}, rings have been proposed for Chiron but consensus is lacking on whether what has been observed was rings, an active jet, or a shell of debris leftover from past activity \cite{Elliot1995,Bus1996,Ruprecht2015,Sickafoose2020}. The most recent results suggest that there is actively-evolving material around Chiron, as opposed to a Chariklo-like ring system \cite{Sickafoose2023,Ortiz2023}. In terms of characterization, occultation data in 1994 were taken in visible wavelengths in ground-based data and open and \textit{K}-band from NASA's Kuiper Airborne Observatory (KAO) and used to place a lower limit of $0.25 \, {\rm \mu m}$ on the radius of particles in the deepest occultation feature \cite{Elliot1995}. This result was noted to be different from comet Halley, for which analyses of spacecraft data found a significant population of particles smaller than this size \cite{Elliot1995}. In 2011, an occultation by Chiron was observed simultaneously in visible ($\sim0.6 \, {\rm \mu m}$) and near-infrared wavelengths (low-resolution spectra from $1-2.5 \, {\rm \mu m}$): the SNR of the near-infrared data was insufficient to detect any surrounding material or color-wavelength trends \cite{Sickafoose2020}. 

Echeclus and SW1 are known to have had significant surrounding material (see Chapters 8 \& 10). Echeclus had a 7-magnitude increase in brightness in 2005 that included a detached coma and substantial dust production \cite{Choi2006,Rousselot2008, Bauer2008}. From \textit{Spitzer} and visible-wavelength data, the maximum size of the ejected particles was estimated to be $700 \, {\rm \mu m}$ within an order of magnitude \cite{Bauer2008}. Maximum activity corresponded to a dust production rate of a few hundreds of kilograms per second, of the order of 30 times that seen in other Centaurs \cite{Rousselot2016}. Impacts and fragmentation were ruled out, but possibly large chunks of material were removed by erosion to expose a fresh surface \cite{Bauer2008}. After a 2017 outburst, visible images plus near-infrared spectra were used to detect a large cloud of ejected debris, leading to speculation that there may be several debris ejection or fragmentation events per year on other Centaurs that are going unnoticed \cite{Kareta2019}.

SW1 is extremely active, with a continuous dust coma and displaying at least one major outburst each year since 1927 \cite{Fink2021}. For example, from 2002-2007, the average rate was 7.3 events per year, and visible-wavelength images showed that the coma was being continuously supplied with fine dust while grains larger than $\sim1\,{\rm \mu m}$ were few in number \cite{Trigo-Rodrguez2008}. A debris trail was detected in \textit{Spitzer} $24 \, {\rm \mu m}$ data in 2003, with an upper limit placed on optical depth \cite{Stansberry2004}. More recently, a 1000-km debris cloud was proposed \citep{Miles2018}, but substantial debris was ruled out within 500-1000 km of the nucleus from a stellar appulse in 2020 \citep{Miles2021}. 

More data are needed to detect and constrain the properties (and evolution) of large debris around Chiron, Echeclus, SW1, and other Centaurs.

\section{Formation Mechanisms}\label{sec:formation}
In contemplating the formation mechanisms for satellites, rings, and debris, an initial question is whether the formation of these components is related. Possible interrelationships include simultaneous formation of debris and a satellite, debris as a source for ring particles, and a satellite as a source for ring particles. Within the Centaur-like population, there are no objects known with both satellites and rings; on the other-hand, both TNOs known to have rings also have satellites. There is not yet enough data to tell whether ring formation is directly related to satellite formation, or whether debris can eventually coalesce into rings.

\subsection{Satellite Formation}
Satellite formation mechanisms can be inferred from the distribution of binary properties. For example, as seen in Figure \ref{fig:binaries_2}, many binaries are near-equal brightness and widely separated. Such objects have far too much angular momentum to form through collisions, and the collision rate is too rare to easily produce such systems in any case. Instead, the dominant formation mechanism for TNO binaries is direct gravitational collapse of a large cloud of gas and small particles. This process naturally forms the types of binaries seen. The concentration mechanism causing the collapse is hypothesized to be the ``streaming instability" where gas-dust-gas interactions concentrate material to the point where self-gravity causes gravitational collapse. Indeed, the distribution of TNO binary angular momenta is a good match for simulations of clouds formed in the streaming instability, making TNO binaries a dominant observational constraint on this mode of planetesimal formation \cite{Nesvorny2020}. 

The streaming instability is particularly effective in the Cold Classical region of the disk. The geomorphology of the TNO Arrokoth, a contact-binary in this region, was recently found to be consistent with a merger of similarly-sized planetesmials from a collapse cloud \cite{Stern2023}. The Hot population of TNOs (and progenitors of most Centaurs and GPCs) is thought to originate in a primordial disk outside a compact configuration of the giant planets (see Chapter 2). The parameters for the streaming instability and binary formation in the primordial disk are poorly constrained. Still, the formation of Centaur-like binaries traces back to their formation as Hot TNOs. Unlike for comet-like activity and debris, the orbital dynamics of Centaur-like objects are not likely to contribute to satellite formation, but rather serve as a destruction mechanism as discussed in Section \ref{sec:sats}. 

\subsection{Ring Formation}\label{subsection:ringformation} 

Soon after the discovery of Chariklo's rings, several studies addressed the question of their formation. The planet-crossing nature of Centaur orbits prompted Smoothed-Particle Hydrodynamics (SPH) simulations to investigate the possibility that the rings, and also small satellites, could be formed by tidal disruption of a differentiated Chariklo during the closest encounter with a giant planet \cite{Hyodo2016}. Such an encounter would remove material from the surface and create a surrounding disk, beyond the Roche limit of which particles could accrete into moons. This model is consistent with the presence of water ice in the rings, as suggested by \cite{Duffard2014}. However, in order for the proposed disruption to occur, the distance between Chariklo and one of the giant planets has to be within a few planetary radii -- much smaller than the likely approach distances \cite{Araujo2016,Wood2017}. In fact, the majority of simulated close encounters  ($\>90\%$ \cite{Araujo2016} or $\>99\%$ \cite{Wood2017}) do not even affect the stability of Chariklo-like rings, making tidal disruption an unlikely formation mechanism.

Rings could be formed during a close encounter with a giant planet that perturbed a small moon inward. This scenario would require a single Neptune encounter and a close match between the satellite orbit and the encounter strength -- if true, it weakly favors formation during the encounter, as opposed to within the Trans-Neptunian region, it would require that most $\sim$100~km TNOs have ten-km sized moons, and it predicts that only a few percent of Centaurs would have rings \cite{Pan2016}. If this scenario is prevalent, rings should be common among large Centaurs but absent for small comets and large TNOs \cite{Pan2016}. Along the same lines, rings might be formed from the disruption of a satellite after crossing the Roche limit: for Chariklo, a porous satellite with a radius of roughly 7~km would disaggregate at the ring location with sufficient mass to explain the rings as well as putative shepherd satellites \cite{Melita2017}. However, this model requires a mechanism to bring the satellite closer to Chariklo, as tidal effects are very weak.

Alternatively, the rings could be created by dusty outgassing or a collision. Chariklo's transition to the Centaur region would have increased its temperature, lofting dust particles off the surface. For particles that were not reaccreted, mutual collisions would settle them into an equatorial ring and frequent collisions could eventually convert dust into meter-sized ring particles \cite{Pan2016}. Collisional considerations include either cratering ejecta or the remnants of a destroyed satellite; however, assuming typical impact probabilities, the estimated timescales for these scenarios are longer than the dynamical lifetime of Centaurs \cite{Melita2017}.

If rings were common around TNOs, they would survive during the transition into the Centaur region \cite{Wood2017}, and formation mechanisms in the more distant Solar System need to be considered. This is an open area of research, with rings only proposed around the TNOs Haumea and Quaoar. Haumea is a unique case, with a collisional history \citep[e.g. ][]{Brown2007} and as a highly-elongated, quickly-rotating body for which rotational fission might have stripped off material \cite{Ortiz2017}. N-body simulations of ring formation from fission at Haumea found that the ring likely formed between an unstable region of the orbit and the Roche radius, near the 3:1 SOR \cite{Sumida2020}. It has been suggested that ring material at Quaoar is the result of a primordial collisional system that settled into a disk, from which both rings and Weywot formed \cite{Morgado2023}. Due to the size difference between TNOs with rings and Centaurs with rings, it is possible that ring formation mechanisms are not directly related. The discovery (or clear absence) of rings around smaller TNOs would be essential for understanding whether rings form in some (likely) primordial process and then survive the transition to Centaur-like orbits, or whether the transition itself generates new rings. 

\subsection{Debris Formation}
If active jets are the source of large debris, then numerous cometary dust models are available to characterize the transient material (e.g. \cite{Tenishev2011}). The larger Centaurs, however, may have enough mass for some of the debris from activity and outbursts to end up on ballistic paths or become orbital \cite{Fink2021}.  While a model such as \cite{Fink2021} cannot be used to understand morphology, it can be used to predict the size scale and density of orbiting particles: for instance, SW1 could have orbiting particles ranging from 8-150 mm that were lifted during outburst events. To understand how these orbiting particles might accrete to form larger bodies, shells, arcs, or rings, additional dynamical modeling is needed.

Although collision rates are likely to be low in the Centaur region, debris could be launched by collisions. Photometric observations of Chiron in 2014-2015 indicated microactivity that was thought to possibly be caused by existing debris impacting the surface, producing outbursts, and launching dust \cite{Cikota2018}.

\section{Summary}\label{sec:conclusions}
Centaur-like bodies are located in a transitional dynamical region. There are many more TNO binaries than Centaur binaries, which could be due to the pathway from dynamically hot populations in the outer Solar System. However, there may be many close binaries which are yet undetected. Ring systems have been proposed for two Centaurs and two TNOs, but observations are few and have mixed interpretations. Confirmation and characterization of stably-orbiting material over longer baselines is needed for all of these objects. Reliable, bulk statistics for satellites and rings for both Centaurs and TNOs will substantially aid understanding of formation mechanisms and locations. Debris has been detected around Centaurs but not at TNOs, suggesting that heliocentric distance plays a role. Importantly, observational biases mean that smaller and fainter features in counterpart TNOs may not yet have been detected.

While we expect that there will be ties between satellites, rings, and debris for bodies in the outer Solar System, we currently lack sufficient information to make concrete connections. The future prospects discussed below (and in Chapter 14) should help answer the many open questions in these fields.

\subsection{Open Questions}\label{subsec:questions}

Here we present a selection of open questions in our current understanding of satellites, rings, and debris around Centaur-like bodies. 

\begin{itemize}

\item \textit{What are the binary fractions of Centaurs and other related dynamical groupings?} Analyses of existing data are needed to determine the unbiased statistics of binary objects, the proportions of tight versus wide orbits, and the percentages of binaries as a function of object size.

\item \textit{What are the shapes and orientations of the known GPC binaries Ceto-Phorcys and Typhon-Echidna?} Even a small amount of new high-precision astrometry would likely detect non-Keplerian effects that would give better insight into the physical properties of these binaries. 

\item \textit{Are there any Centaur ``pairs" that indicate a previously-bound binary system?} Even with a lower limit of $\sim$10\% binary fraction of Hot TNOs, if the current, low frequency of Centaur-like binaries is attributable to the unbinding of TNO binaries in close encounters, then there would be a significant number of Centaur ``pairs'': objects with orbits so similar that they must have been physically close in the past. These same close encounters cause the orbits of Centaurs to be chaotic, making it challenging to reliably trace orbits back far in time. Thus, Centaur pairs from binary formation would require a recent interaction. It is not clear whether the currently-known population of hundreds of Centaur-like objects is sufficient to detect any ``pairs".

\item \textit{What are the effects on binary systems of outbursting activity or binary-binary interactions?} These factors could play roles in the apparent drop in binary fraction from TNOs to Centaurs and JFCs; however, binary-binary interactions are likely to be relevant only for very close encounters.

\item \textit{How can light-curve observations be better employed to learned about Centaur and TNO satellites?} Light curves provide an indirect method of detecting contact binaries and there are indications that light-curve amplitudes differ between dynamical populations. However, Centaur outbursting could be repressing amplitudes and there remain observational biases.
    
\item \textit{How (and where) do Centaur rings form?} A major factor distinguishing Centaur ring-formation theories is whether (i) these rings are primordial and formed in the TNO region and were maintained throughout the dynamical transition to Centaurs, or (ii) the Centaur-precursor, small, Hot TNOs do not have rings, requiring either planetary encounters or heliocentric-distance-induced surface activity to generate them. It is unlikely that we happen to have detected rings during a short period of time when they exist, rather they are likely long-term features and may even be prevalent among Centaurs. The frequency of rings, especially with well-determined upper limits, is still poorly understood, especially for small TNOs of similar sizes to ringed Centaurs. 
 
\item \textit{What is the orbital evolution of material around active Centaurs?} More work is needed to understand the dynamics of larger-than-dust particles around active Centaurs, whether debris can coalesce into rings, and whether rings can be maintained or will disperse over relatively short timescales. A subset of this question is the following:\textit{ how is orbiting material evolving at Chiron?} Observations are specifically needed to estimate the particle sizes and locations of current debris at Chiron, as well as to continue to study its evolution.

\item \textit{Are there any promising Centaur targets for as-yet-undetected ring systems?} The spin-orbit coupling theory suggests that faster rotators may be promising targets. Occultations and appulses by the Centaur Bienor (9.14-hr rotation) have accordingly been targeted, but Chariklo-like rings could not be ruled out \cite{Fernandez-Valenzuela2023}. The limited number of known small-body ring systems makes it difficult to isolate ring-friendly characteristics. 
    
\end{itemize}

\subsection{Future Prospects}

There are many possibilities for enhanced understanding of binaries, rings, and debris using existing datasets or observations expected in the near-term. Several of the open questions above can be addressed with current data, though anticipated discoveries in the next few years will significantly augment research opportunities. 

With existing data, a clear step forward is better characterization of non-detections. For example, there are no detailed publications of non-detections or upper limits for binary components around the 109 Centaur-like objects observed with HST (or for the few hundred TNOs observed with HST). This makes it difficult to determine whether the binary frequency among similarly-sized objects is truly distinct. 

Similarly, occultation observations of rings and debris suffer from a completeness problem. The occultation field is transitioning from isolated, anecdotal detections of individual objects to understanding statistics of multiple objects which will require a new threshold of expectations. For example, published occultation observations do not always provide all the information necessary for homogeneous reanalysis of data, and non-detections or upper limits often remain unpublished. Although reporting null results and carefully placing upper limits is a challenging prospect for occultation observers, it is an important goal going forward. 

Chapter 14 discusses new surveys and instruments expected to enhance Centaur science. Below, we briefly discuss how these would specifically be applied to Centaur satellites, rings, and debris. 

\subsubsection{JWST}
JWST observations of Centaurs are already ongoing, but they are expected to be generally limited to only the most interesting objects. 

In its shortest wavelength filter, JWST's NIRCam can detect binaries closer and fainter than HST, but only by a factor of $\sim$2, so this is not expected to be a significant source of new binaries. Still, some high precision astrometry from JWST may complement continued observations with HST to learn more about binary fractions and orbits. 

JWST has been effectively used to detect Chariklo's rings during a stellar occultation \cite{Santos-Sanz2023}. It is also possible that JWST high-resolution imaging could resolve rings of Centaurs or TNOs. Direct imaging of the rings would provide new and unique information about their full orbital structure, composition, and temporal evolution. 

\subsubsection{Vera C. Rubin Observatory Legacy Survey of Space and Time (LSST)}

The upcoming LSST is expected to increase the number of known small bodies throughout the Solar System by an order of magnitude. Furthermore, there are typically hundreds of detections processed to produce highly-precise astrometry and photometry in $ugriz$ filters. Thus, the quality of small-body observations will also increase by more than an order of magnitude. The vast majority of small-body discoveries by LSST will occur in the first year or two of the survey, suggesting that the number of known Centaur-like objects will increase substantially in 2026-2027 (based on present estimates for the LSST schedule). 

With both accurate (thanks to GAIA) and precise astrometry, the orbits of Centaurs will be improved significantly, allowing for increased success in occultation surveys. This improvement will be plausible within the first year of the LSST survey for established objects, since new astrometry will significantly extend observational baselines and improve known positions. For new discoveries, such precise orbits are expected to take at least two years, but they may be possible for occultation predictions even earlier. 

LSST's precise, multi-epoch, and multi-filter photometry will also enable detailed studies of practically all Centaur-like objects that are usually reserved for those with ground-based observing campaigns. Thousands of Centaur-like objects should have well-measured rotational amplitudes and periods, as well as colors. The rotational-period distribution then may give insights into binary frequency (as in Section \ref{subsec:otherbin}), while amplitudes will help constrain contact-binary fractions. Even low-level outbursting activity will be detectable at unprecedented levels with LSST, allowing for better characterization of surrounding material and understanding of possible debris-forming events. 

\subsubsection{Extremely Large Telescopes}
Future extremely large telescopes open the possibility of significant advances in characterizing the near-Centaur environment. The Extremely Large Telescope in Chile will have direct imaging capabilities with 5 mas resolution with an intensity contrast of better than $10^{-8}$ at 30 mas \cite{2010SPIE.7735E..2EK}. This capability will enable a systematic search of Centaurs for satellites, rings and debris. As described in the TMT Detailed Science Case 2022\footnote{Available at https://www.tmt.org/documents.}, the Thirty Meter Telescope's IRIS instrument provides diffraction limited imaging and will be able to resolve ring structures similar to Chariklo's with its 7 mas spatial resolution at 1 $\mu$m.

\subsubsection{Potential Spacecraft Missions}

The Centaurs are arguably the main dynamical category of small bodies in the Solar System that have not yet had the benefit of a close study enabled by a spacecraft mission. A mission to a Centaur is more practical than a visit to a distant TNO and could answer many similar questions. Further discussion is in Chapter 16, but we note here the special value of choosing a mission target that is a binary, has rings, or is known to have had debris. The potential insights into these phenomena from a close flyby are extensive and compelling. 

\section{Epilogue}
Centaurs (including Giant Planet Crossers) are an interesting population themselves and further serve as a unique interface between the TNO and JFC populations. The properties of binaries, rings, and debris provide insights into the physical, dynamical, and evolutionary properties of these objects. Observations -- both for individual objects and in terms of homogeneously-analyzed surveys -- are quite limited and currently leave many questions open and unclear. There are ample prospects for future analyses of these features, especially with the anticipated improvements in quantity and quality of observations from JWST, LSST, and extremely large telescopes. Further investigations will provide powerful insights into the formation and evolution of Centaurs as well as the outer Solar System in general. 

\bibliographystyle{unsrt}
\bibliography{Bibliography}

\end{document}